\documentclass[a4paper,useAMS,usenatbib]{mn2e}
\usepackage{graphicx}
\usepackage{natbib}
\usepackage{amsmath}
\usepackage{amssymb}
\usepackage{mathrsfs}
\usepackage[usenames]{color}
\usepackage{hyperref}
\usepackage[normalem]{ulem} 
\usepackage[section]{placeins}
\usepackage{multirow}
\usepackage{enumerate}

\newcommand{\feh}{\mbox{\rm [{\rm Fe}/{\rm H}]}}
\newcommand{\mh}{\mbox{\rm [{\rm M}/{\rm H}]}}
\newcommand{\Msun}{\mbox{$M_{\odot}$}}
\newcommand{\Teff}{\mbox{$T_{\rm eff}$}}

\newcommand{\beq}{\begin{equation}}
\newcommand{\eeq}{\end{equation}}
\newcommand{\beqa}{\begin{eqnarray}}
\newcommand{\eeqa}{\end{eqnarray}}
\newcommand{\deltanu}{\mbox{$\Delta\nu$}}
\newcommand{\numax}{\mbox{$\nu_{\rm max}$}}
\newcommand{\deltaP}{\mbox{$\Delta P$}}

\newcommand{\comment}[1]{}

\newcommand{\white}{\textcolor{white}}
\newcommand{\aovh}{\mbox{$\alpha_{\rm ovH}$}}
\newcommand{\aovhe}{\mbox{$\alpha_{\rm ovHe}$}}
\newcommand{\brunt}{Brunt-V\"ais\"al\"a}
\title[Beyond scaling relations]{Determining stellar parameters of asteroseismic targets: going beyond the use of scaling relations}

\author[Rodrigues et al.]{%
Tha\'ise S. Rodrigues$^{1,2}$, 
Diego Bossini$^{3}$, 
Andrea Miglio$^{3,4}$,
L\'eo Girardi$^{1}$, \newauthor  
Josefina Montalb\'an$^{2}$, 
Arlette Noels$^{5}$,  
Michele Trabucchi$^{2}$,\newauthor
Hugo Rodrigues Coelho$^{3,4}$, 
Paola Marigo$^{2}$
  \\ 
  $^{1}$ Osservatorio Astronomico di Padova -- INAF, Vicolo dell'Osservatorio 5, I-35122 Padova, Italy \\ 
  $^{2}$ Dipartimento di Fisica e Astronomia, Universit\`a di Padova,  Vicolo dell'Osservatorio 2, I-35122 Padova, Italy \\
  $^{3}$School of Physics and Astronomy, University of Birmingham, Edgbaston, Birmingham B15 2TT, United Kingdom \\
  $^{4}$Stellar Astrophysics Centre, Department of Physics and Astronomy, Aarhus University,  DK-8000 Aarhus C, Denmark \\
  $^{5}$ Institut d'Astrophysique et de G\'eophysique, All\'ee du 6 aout, 17 -- Bat.\ B5c, B-4000 Li\`ege 1 (Sart-Tilman), Belgium }

\begin{document}

\date{Accepted 2017 January 13 for publication in MNRAS}


\maketitle

\label{firstpage}

\begin{abstract}
Asteroseismic parameters allow us to measure the basic stellar properties of field giants observed far across the Galaxy. Most of such determinations are, up to now, based on simple scaling relations involving the large frequency separation, \deltanu, and the frequency of maximum power, \numax. In this work, we implement \deltanu\ and the period spacing, \deltaP, computed along detailed grids of stellar evolutionary tracks, into stellar isochrones and hence in a Bayesian method of parameter estimation. Tests with synthetic data reveal that masses and ages can be determined with typical precision of 5 and 19 per cent, respectively, provided precise seismic parameters are available. Adding independent information on the stellar luminosity, these values can decrease down to 3 and 10 per cent respectively. The application of these methods to NGC~6819 giants produces a mean age in agreement with those derived from isochrone fitting, and no evidence of systematic differences between RGB and RC stars. The age dispersion of NGC~6819 stars, however, is larger than expected, with at least part of the spread ascribable to stars that underwent mass-transfer events.
\end{abstract}


\section{Introduction}
\label{intro}
  
With the detection of solar-like oscillation in thousands of red giant stars, {\it Kepler} and CoRoT missions have opened the way to the derivation of basic stellar properties such as mass and age even for single stars located at distances of several kiloparsecs \citep[e.g.][and references therein]{chaplinmiglio13}. In most cases this derivation is based on the two more easily-measured asteroseismic properties: the large frequency separation, \deltanu, and the frequency of maximum oscillation power, \numax. \deltanu\ is the separation between oscillation modes with the same angular degree and consecutive radial orders, and scales to a very good approximation with the square root of the mean density ($ \overline{\rho}$), while \numax\ is related with the cut-off frequency for acoustic waves in a isothermal atmosphere, which scales with surface gravity $g$ and effective temperature \Teff. These dependencies give rise to the so-called scaling relations:
\begin{eqnarray}
\Delta\nu & \propto & \overline{\rho}^{1/2} \propto M^{1/2} / R^{3/2} \nonumber \\ 
\nu_{\rm max} & \propto & g T_{\rm eff}^{-1/2} \propto (M/R^2)  T_{\rm eff}^{-1/2} \,\,\,\,.
\label{eq:scaling}
\end{eqnarray}
It is straightforward to invert these relations and derive masses $M$ and radii $R$ as a function of \numax, \deltanu, and \Teff. The latter has to be estimated in an independent way, for instance via the analysis of high-resolution spectroscopy. $M$ and $R$ can then be determined either (1) in a model-independent way by the ``direct method'', which consists in simply applying the scaling relations with respect to the solar values, or (2) via some statistical method that takes into account stellar theory predictions and other kinds of prior information. In the latter case, the methods are usually referred to as either ``grid-based'' or ``Bayesian'' methods.

Determining the radii and masses of giant stars brings consequences of great astrophysical interest: The radius added to a set of apparent magnitudes can be used to estimate the stellar distance and the foreground extinction. The mass of a giant is generally very close to the turn-off mass of its parent population, and hence closely related with its age; the latter is otherwise very difficult to estimate for isolated field stars.  In addition, the surface gravities of asteroseismic targets can be determined with an accuracy generally much better than allowed by spectroscopy.

Although these ideas are now widely-recognized and largely used in the analyses of CoRoT and {\it Kepler} samples, there are also several indications that asteroseismology can provide even better estimates of masses and ages of red giants, than allowed by the scaling relations above. First, there are significant evidences of corrections of a few percent being necessary \citep[see ][]{white11,
miglio12_1,miglio13,brogaard16, miglio16, guggenberger16, sharma16, handberg16} in the $\deltanu$ scaling relation. Although such corrections are expected to have little impact on the stellar radii (and hence on the distances), they are expected to reduce the errors in the derived stellar masses, hence on the derived ages for giants. Second, there are other asteroseismic parameters as well -- like for instance the period spacing of mixed modes, $\Delta P$ \citep{beck11, mosser14} -- that can be used to estimate stellar parameters, although not via so easy-to-use scaling relations as those above-mentioned.
 
In this paper, we go beyond the use of simple scaling relations in the estimation of stellar properties via Bayesian methods, first by replacing the \deltanu\ scaling relation by using frequencies actually computed along the evolutionary tracks, and second by including the period spacing $\deltaP$ in the method. We study how the precision and accuracy of the inferred stellar properties improve with respect to those derived from scaling relations, and how they depend on the set of available constraints.
The set of additional parameters to be explored includes also the intrinsic stellar luminosity, which will be soon determined for a huge number of stars in the Milky Way thanks to the upcoming Gaia parallaxes \citep[][and references therein]{lindegren16}. The results are tested both on synthetic data and on the star cluster NGC~6819, for which \textit{Kepler} has provided high-quality oscillation spectra for about 50 giants \citep{basu11, stello11_1, corsaro12, handberg16}.

The structure of this paper is as follows. Section~\ref{sec:models} presents the grids of stellar models used in this work, describes how the \deltanu\ and \deltaP\ are computed along the evolutionary tracks, and how the same are accurately interpolated in order to generate isochrones. Section~\ref{sec:applications} employs the isochrone sets incorporating the new asteroseismic properties to evaluate stellar parameters by means of a Bayesian approach. The method is tested both on synthetic data and on real data for the NGC~6819 cluster. Section~\ref{sec:close} draws the final conclusions.

\section{Models}
\label{sec:models}

\subsection{Physical inputs}

The grid of models was computed using the MESA code  \citep{Paxton_etal11,Paxton_etal13}. 
We computed 21 masses in a range between $M= 0.6-2.5 \Msun$, in combination with 7 different metallicities ranging from [Fe/H]$=-1.00$ to $0.50$ (Table \ref{tab:grid})\footnote{According to the simulations by \citet{girardi15}, less than one per cent of the giants in the Kepler fields are expected to have masses larger than 2.5~\Msun.}.
The following points summarize the relevant physical inputs used:
\begin{itemize}
\item The tracks were computed starting from the pre-main sequence (PMS) up to the first thermal pulse of the asymptotic giant branch (TP-AGB).
\item We adopt \citet{GN93} heavy elements partition. 
\item The OPAL equation of state \citep{Rogers&Nayfonov02} and OPAL opacities \citep{iglesias96} were used, augmented by low-temperature opacities from \citet{Ferguson_etal05}. C-O enhanced opacity tables were considered during the helium-core burning (HeCB) phase.
\item A custom table of nuclear reaction rates was used \citep[NACRE,][]{Angulo_etal99}. 
\item The atmosphere is taken according to \citet{Krishna-Swamy66} model. 
\item Convection was treated according to mixing-length theory, using the solar-calibrated parameter ($\alpha_\mathrm{MLT}=1.9657$). 
\item Overshooting was applied during the core-convective burning phases in accordance with \citet{Maeder75} step function scheme. We use overshooting with a parameter of $\aovh=0.2 H_p$ during the main sequence, while we consider $\aovhe=0.5 H_p$ penetrative convection in the HeCB phase     
(following the definitions in \citealt{Zahn91} and the result in \citealt{bossini15}).    
\item Element diffusion, mass loss, and effects of rotational mixing were not taken in account.
\item Metallicities [Fe/H] were converted in mass fractions of heavy elements $Z$ by the approximate formula $Z=Z_{\odot}\cdot10^{\mathrm{[Fe/H]}}$, where $Z_{\odot}=0.01756$, coming from the solar calibration.  The initial helium mass fraction $Y$ depends on $Z$ and was set using a linear helium enrichment expression 
\begin{equation}
Y=Y_p+\frac{\Delta Y}{\Delta Z} Z
\label{eq:Y_enrich}
\end{equation}
with the primordial helium abundance $Y_{p} = 0.2485$ and the slope $\Delta Y/\Delta Z =(Y_\odot-Y_p)/Z_\odot =1.007$. 
Table \ref{tab:grid} shows the relationship between [Fe/H], $Z$, and $Y$ for the tracks computed.
\end{itemize}

\begin{table}
\scriptsize
\centering
\caption{Initial masses and chemical composition of the computed tracks.\label{tab:grid}}
    \begin{tabular}{ c }
    \hline     Mass (M$_\odot$)  \\
    \hline     0.60, 0.80, 1.00, 1.10, 1.20, 1.30, 1.40, 1.50, 1.55, 1.60,\\
               1.65, 1.70, 1.75, 1.80, 2.00, 2.15, 2.30, 2.35, 2.40, 2.45, 2.50 \\
    \hline
    \end{tabular}
    \begin{tabular}{ c  c  c }
    \hline     [Fe/H]  &    $Z$   &    $Y$   \\
    \hline
    $-$1.00  &  0.00176 &  0.25027 \\
    $-$0.75  &  0.00312 &  0.25164 \\
    $-$0.50  &  0.00555 &  0.25409 \\
    $-$0.25  &  0.00987 &  0.25844 \\
    0.00  &  0.01756 &  0.26618 \\
    0.25  &  0.03123 &  0.27994 \\
    0.50  &  0.05553 &  0.30441 \\
    \hline
    \end{tabular}
\end{table}

\subsection{Structure of the grid}
\label{sec:grid}

To build the tracks actually used in our Bayesian-estimation code, we select from the original tracks computed with MESA about two hundred structures well-distributed in the HR diagram and representing all evolutionary stages. From these models we extract global quantities, such as the age, the photospheric luminosity, the effective temperature (\Teff), the period spacing of gravity modes (\deltaP, see Section~\ref{sec:deltaP}).
In addition, each structure is also used to compute individual radial mode frequencies with GYRE \citep{Townsend&Teitler13} in order to calculate large separations (\deltanu), as described in Section~\ref{sec:averageDnu}.  

\subsection{Average large frequency separation}
\label{sec:averageDnu}

\subsubsection{Determination of the large frequency separation}
\label{sec:DetAverageDnu}

\begin{figure}
\resizebox{\hsize}{!}{\includegraphics{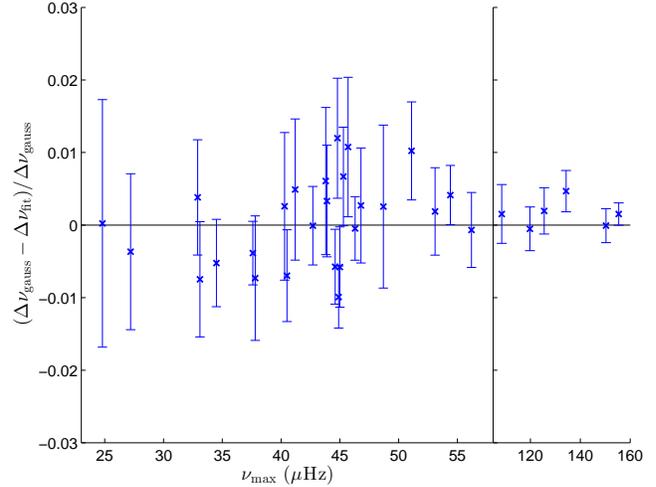}}
    \caption{Comparison between the average large separation $\langle\deltanu_\mathrm{fit}\rangle$ of the star in NGC~6819, estimated by linear fitting with the actual error, and the output of the method described in Section~\ref{sec:DetAverageDnu}, for which the actual errors were substituted by a Gaussian function centred in \numax.}
\label{fig:dnu_methods}
\end{figure}

In a first approximation, the large separation \deltanu\ can be estimated in the models by the equation \ref{eq:scaling}. However, this estimation can be inaccurate, since is affected by systematic effects which depend e.g.\ on the evolutionary phase and, more generally, on how the sound speed behaves in the stellar interior. 
To go beyond the seismic scaling relations, we calculate individual radial-mode frequencies for each of the models in the grid. Based of the frequencies we compute an average large frequency separation $\langle\deltanu\rangle$. We adopt a definition of $\langle\deltanu\rangle$ as close as possible to the observational counterpart. 
The average \deltanu\ as measured in the observations depends on  the number of frequencies identified around \numax\ and on their uncertainties.
Therefore, with the aim of a self-consistent comparison between data and models, any $\langle\deltanu\rangle$ calculated from stellar oscillation codes must take in account the restrictions given by the observations. \citet{handberg16} estimated the quantity $\Delta\nu_\mathrm{fit}$ for the stars in the {\it Kepler}'s cluster NGC~6819. In that paper,  $\Delta\nu_\mathrm{fit}$ is estimated by a simple linear fit of the individual frequencies (weighted on their errors) as function of the radial order. The value of the slope resulting from the fitting line gives the estimated \deltanu.
However, the same method cannot be applied to theoretical models since their frequencies have no error bars. Therefore we need to take into account the uncertainties associated to each frequency in order to give them a consistent weight. 
Observational errors depend primarily on the frequency distance between a given oscillation mode and $\nu_\mathrm{max}$, with a trend that follows approximately the inverse of a Gaussian envelope \citep[smaller errors near \numax, larger errors far away from \numax;][]{handberg16}.
For this reason we adopt a Gaussian function, as described in \citet{Mosser12}, to calculate the individual weights:
\begin{equation}
w=\exp\left[-\frac{(\nu-\nu_\mathrm{max})^2}{2\cdot\sigma^2}\right],
\label{eq:gauss}
\end{equation}
where $w$ is the weight associated to the oscillation frequency $\nu$, and 
\begin{equation}
\sigma=0.66\cdot\nu_\mathrm{max}^{0.88} \,\,\,\,. 
\label{eq:sigma_envelope}
\end{equation}
The $\langle\deltanu\rangle$ is then calculated by a linear fitting of the radial frequencies $\nu_{n,0}$ as function of the radial order $n$, with the weights taken at each $\nu_{n,0}$ frequency. 
In order to test our estimations we use the observed frequencies in \citet{handberg16} simulating their errors using the Gaussian weight function in equation~\ref{eq:gauss}. 
Figure \ref{fig:dnu_methods} shows the comparison between $\langle\deltanu_\mathrm{gauss}\rangle$, determined from the method above, with $\langle\deltanu_\mathrm{fit}\rangle$ estimated in the paper using the actual errors.
The method   estimate $\langle\deltanu_\mathrm{gauss}\rangle$ with relative differences within the error bars for the majority of the stars.  
Although the definition of $\langle\deltanu\rangle$ may seem a minor technical issue, it plays an important role in avoiding systematic effects on e.g. the mass and age estimates.
 
\subsubsection{Surface effects}

It is well known that current stellar models suffer from an inaccurate description of near-surface layers leading to a mismatch between theoretically predicted and observed oscillation frequencies. These so-called surface effects have a sizable impact also on the large frequency separation, and on its average value. When using model-predicted $\deltanu$ it is therefore necessary to correct for such effects. As usually done, a first attempt at correcting is to use the Sun as a reference, hence by normalising the $\langle\deltanu\rangle$ of a solar-calibrated model with the observed one.

\begin{figure}
    \includegraphics[height=\hsize, angle=-90]{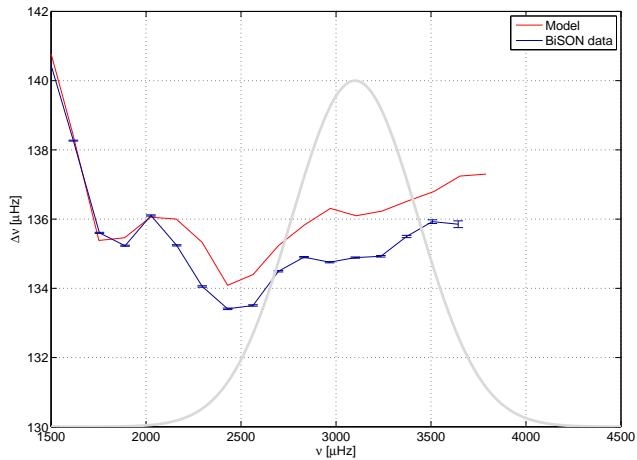}
    \caption{Large frequency separation (\deltanu) of radial modes as function of frequency, as observed in the Sun (\citealt{broomhall14}, dots connected by a blue line) and in our calibrated solar model (red line). The gray gaussian profile represents the weights given by each point of \deltanu\ when estimating $\langle\deltanu\rangle$ (accordingly to the method described in Section~\ref{sec:DetAverageDnu}).}
    \label{fig:dnusun}
\end{figure}

In our solar model, $\alpha_{\rm MLT}$ and $X_{\odot}$ are calibrated  
to reproduce, at the solar age $t_\odot=4.57$ Gyr, the observed luminosity $L_\odot = 3.8418\cdot10^{33}$ erg s$^{-1}$, the photospheric radius $R_\odot = 6.9598\cdot10^{10}$ cm \citep{Bahcall_etal05}, and the present-day ratio of heavy elements to hydrogen in the photosphere ($Z/X=0.02452$, \citealt{GN93}). We used the same input physics as described in Section~\ref{sec:models}.
A comparison between the large frequency separation of our calibrated solar model and that from solar oscillation frequencies \citep{broomhall14} is shown in Fig.~\ref{fig:dnusun}. We find that the predicted average large separation, $\langle\deltanu_{\odot, \rm mod}\rangle=136.1$~$\mu$Hz (defined cf.~Section~\ref{sec:averageDnu}), is 0.8 per cent larger than the observed one ($\langle\deltanu_{\odot, \rm obs}\rangle=135.0$~$\mu$Hz). 
We then follow the approach by \citet{white11} and adopt as a solar reference value that of our calibrated solar model ($\langle\deltanu\rangle_\odot=\langle\deltanu\rangle_{\rm mod, \odot}=136.1 \mu$Hz).

This is an approximation which should be kept in mind, and an increased accuracy when using $\langle\deltanu\rangle$ can only be achieved by both improving our theoretical understanding of surface effects in stars other than the Sun \citep[e.g. see][]{sonoi15, ball16}, and by trying to mitigate surface effects when comparing models and observations.
In this respect a way forward would be to determine the star's mean density by using the full set of observed acoustic modes, not just their average frequency spacing. 
This approach was carried out in at least two RGB stars \citep{huber13, Lillo-Box2014}, and led to determination of the stellar mean density which is $\sim 5-6$ per cent higher than derived from assuming scaling relations, and  with a much improved precision of $\sim 1.4$ per cent. 
Furthermore, the impact of surface effects on the inferred mean density is mitigated when determining the mean density using individual mode frequencies rather than using the average large separation \citep[e.g., see][]{chaplinmiglio13}.
This approach is however not yet feasible for populations studies, mostly because individual mode frequencies are not available yet for such large ensembles, but it is a path worth pursuing to improve both precision and accuracy of estimates of the stellar mean density.


\subsubsection{$\Delta\nu$: deviations from simple scaling} 

Small-scale deviations from the $\langle\deltanu\rangle$ scaling relation  have been investigated in several papers. This is usually done by comparing how well model predicted $\langle\deltanu\rangle$ scales with $\overline{\rho}^{1/2}$, taking the Sun as a reference point \citep[see ][]{white11,
miglio12_1,miglio13,brogaard16, miglio16, guggenberger16, sharma16, handberg16}. 

Such deviations may be expected primarily for two reasons.  First, stars in general are not homologous to the Sun, hence the sound speed in their interior (hence the total acoustic travel time) does not simply scale with mass and radius only. Second, the oscillation modes detected in stars do not adhere to the asymptotic approximation to the same degree as in the Sun  \citep[see e.g.][for a more detailed explanation]{Belkacem2013}.

The combination of these two factors is what eventually determines a deviation from the scaling relation itself. Cases where a small correction is expected  
are likely the result of a fortuitous cancellation of the two effects (e.g. in RC stars).

We would like to stress that beyond global trends e.g. with global properties, such  corrections are also expected to be evolutionary-state and mass dependent, as discussed e.g.\ in \citet{miglio12_1}, \citet{miglio13}, and   \citet{christensendalsgaard14}.  As pointed out in these papers, the mass distribution is very different inside stars with same mass and radius but in RGB or RC phases. A RGB model has a central density $\sim\!10$ times higher than a RC one; the former has a radiative degenerate core of He, while the latter has a very small convective core inside a  He-core. The mass coordinate of the He-core is roughly a factor 2 larger for the RC model, while the fractional radius of this core is very small ($\sim 2.5-6 \times 10^{-3}$) in both cases. The frequencies of radial modes are dominated by envelope properties, which have not very different temperatures. How the difference in the deep interior of the star affects then the relation between mean density and the seismic parameter? As suggested in the above mentioned papers, different distribution of mass implies a lower density of the envelope of the RC with respect to the RGB one, and hence a different sound speed in the regions effectively probed by radial oscillations.  As shown by \citet[][and references therein]{ledoux58}, the oscillation frequencies of radial modes depend not only on the mean density of the star, but also on the mass concentration, with mode frequencies (and hence separations) increasing with  mass concentration. Although in the RGB model the center density is 10 times larger than the same in the RC one, the latter is a more concentrated model  since for 1 M$_{\odot}$, for instance,  half of the stellar mass is inside some thousandths of its radius. Moreover, as mass concentration increases, the oscillation modes tend to propagate in more external layers. Hence, not only the envelope of the RC model has a lower density, in addition the eigenfunctions propagate in more external regions with respect to their behaviour in RGB stars. The adiabatic sound speed of the regions probed by these oscillation modes is smaller in the RC than in the RGB, leading to differences in large frequency separations,  and corrections with respect to the scaling relation. 

\begin{figure*}
    \begin{minipage}{0.4\textwidth}
    \centering
    \resizebox{0.9\textwidth}{!}{\includegraphics{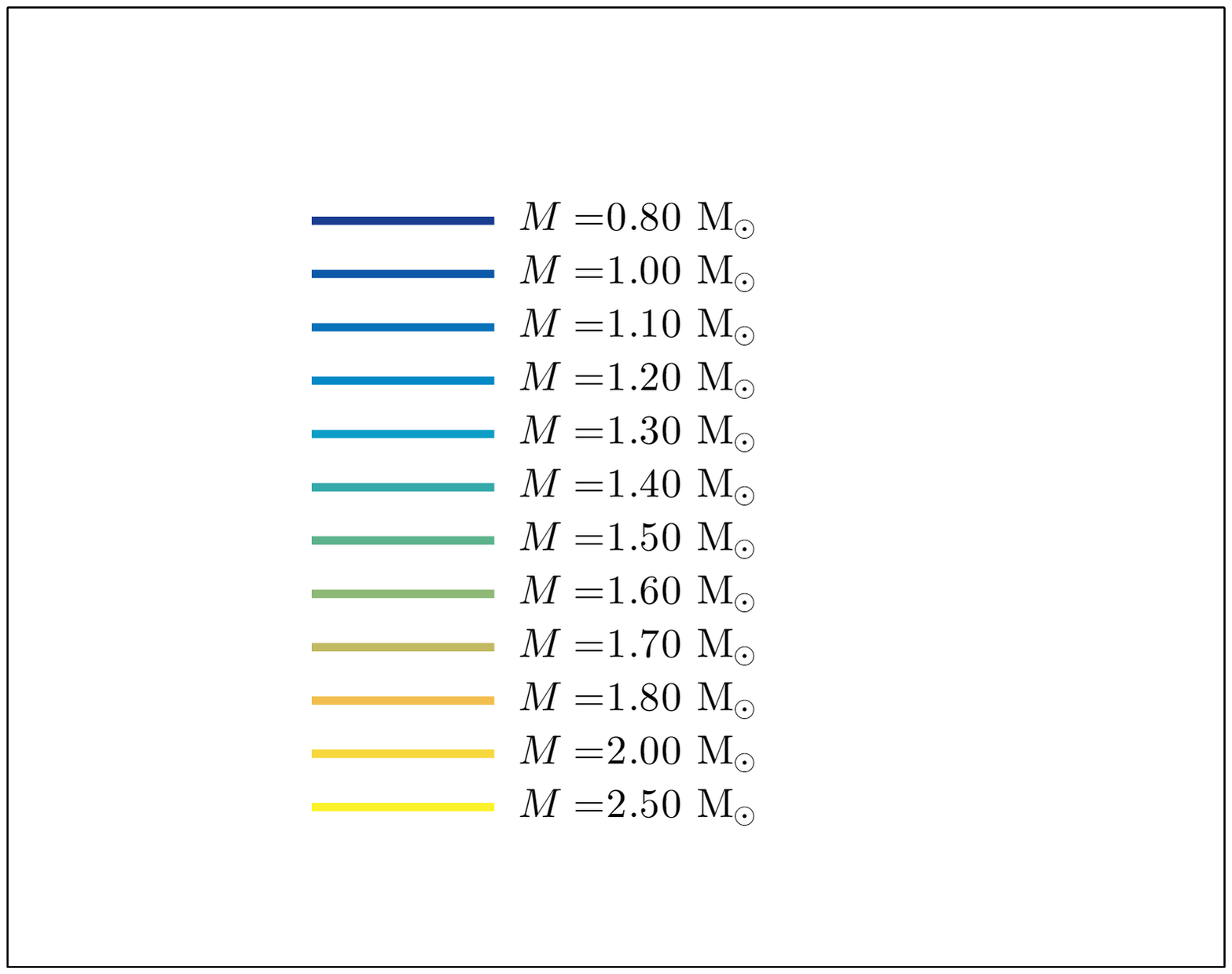}}
    \white{\bf{\\0\\}}
    \resizebox{0.9\textwidth}{!}{\includegraphics{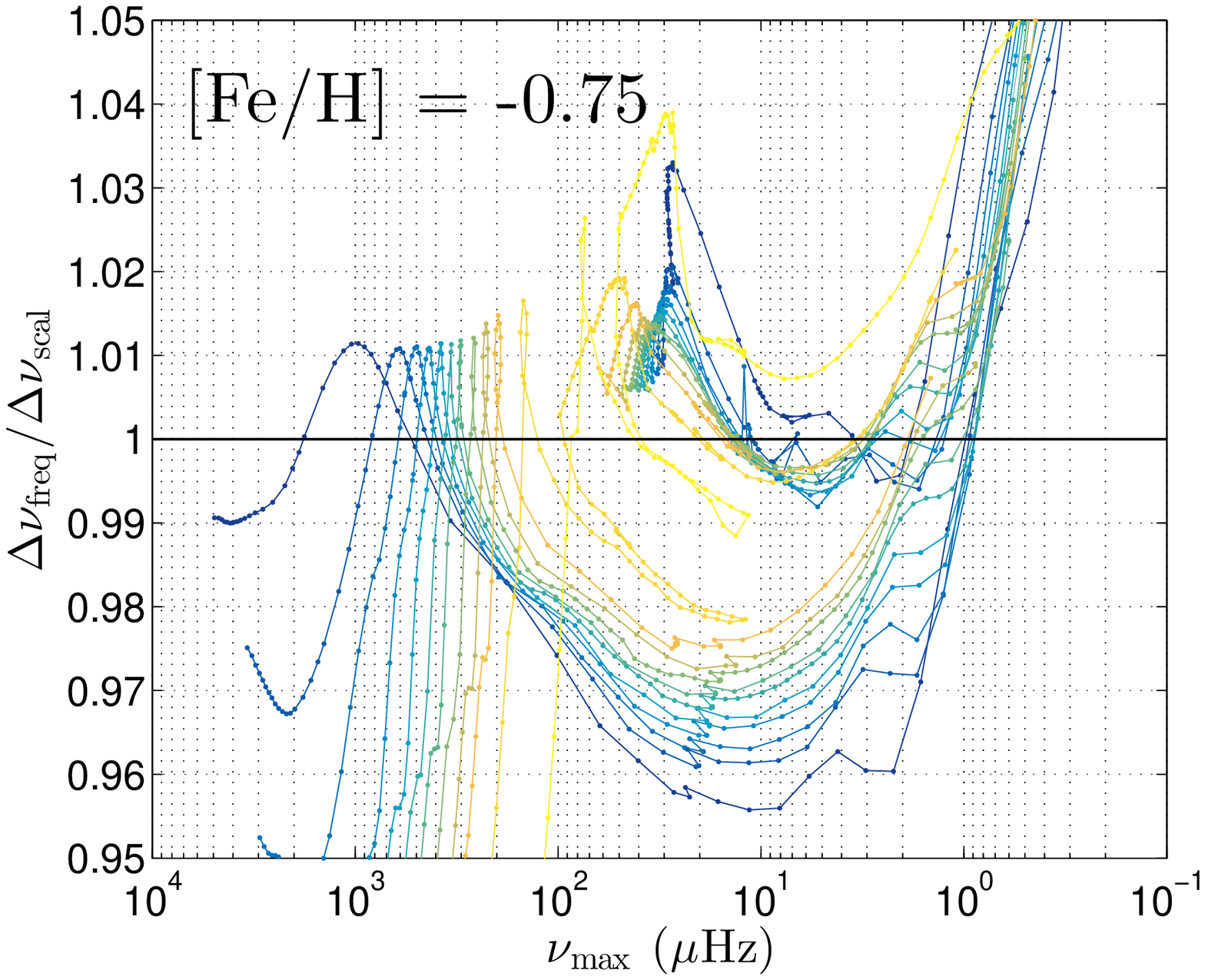}}
    \white{\bf{\\0\\}}
    \resizebox{0.9\textwidth}{!}{\includegraphics{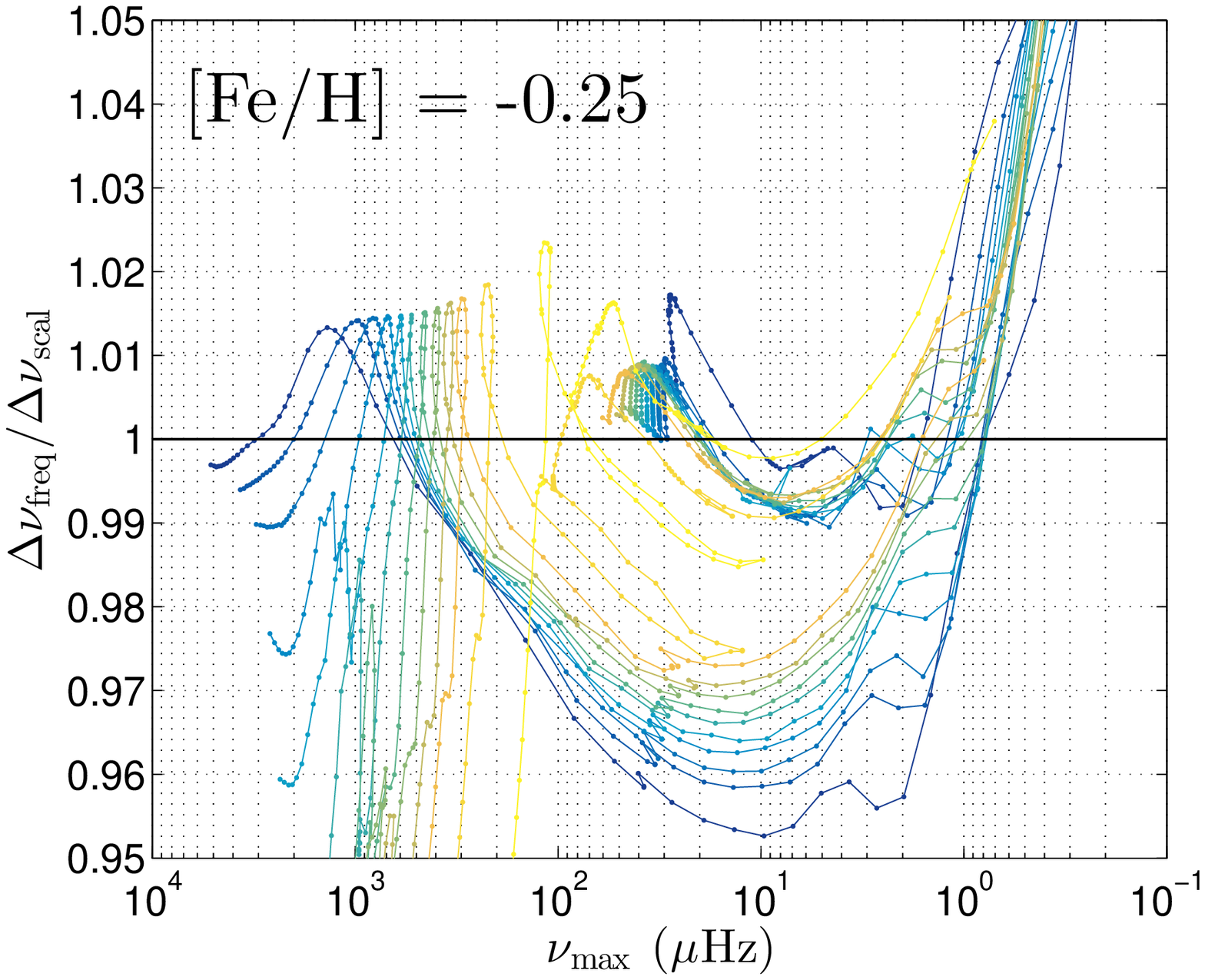}}
    \white{\bf{\\0\\}}
    \resizebox{0.9\textwidth}{!}{\includegraphics{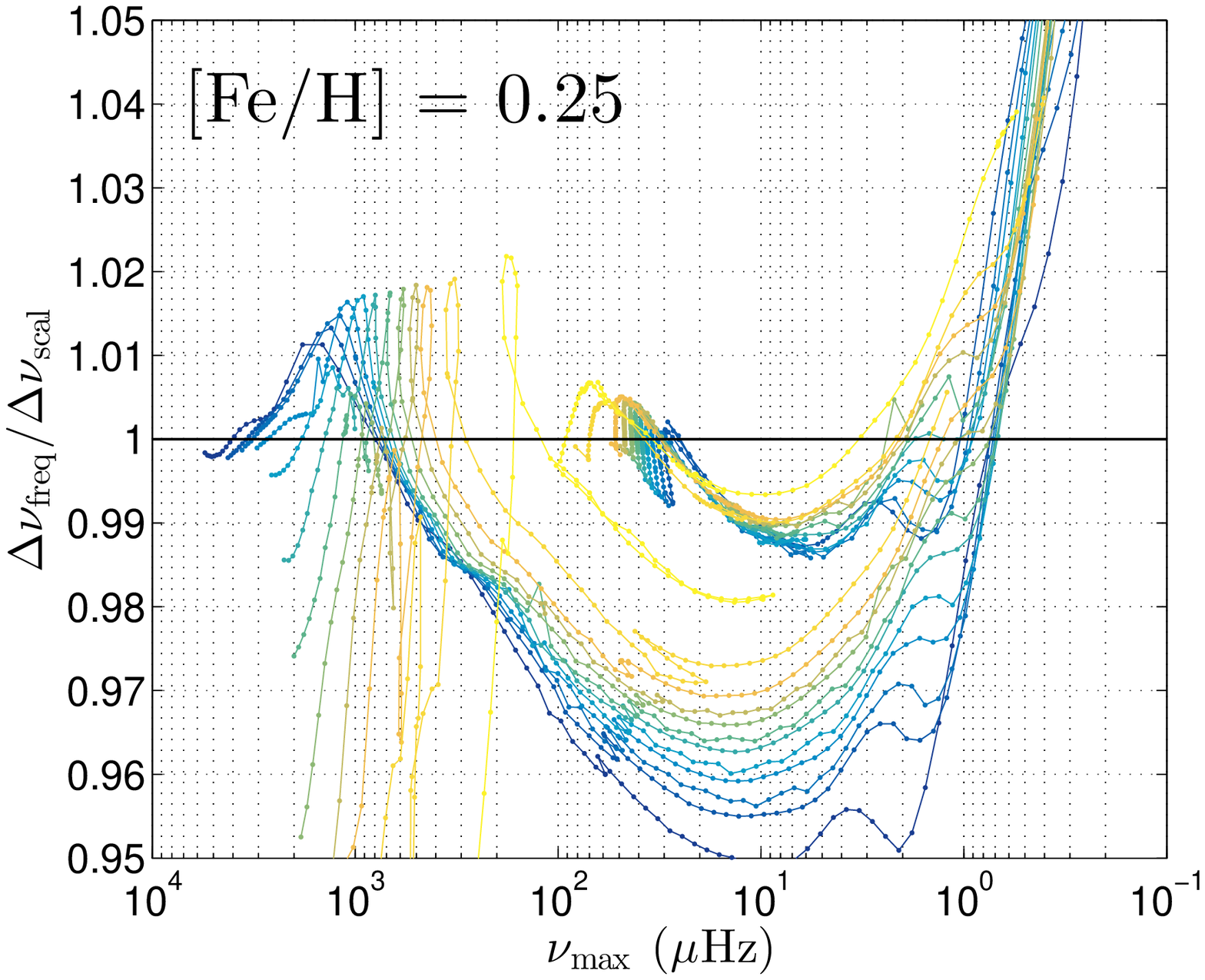}}
    \white{\bf{\\0\\}}
    \end{minipage}
    \begin{minipage}{0.4\textwidth}
    \centering
    \resizebox{0.9\textwidth}{!}{\includegraphics{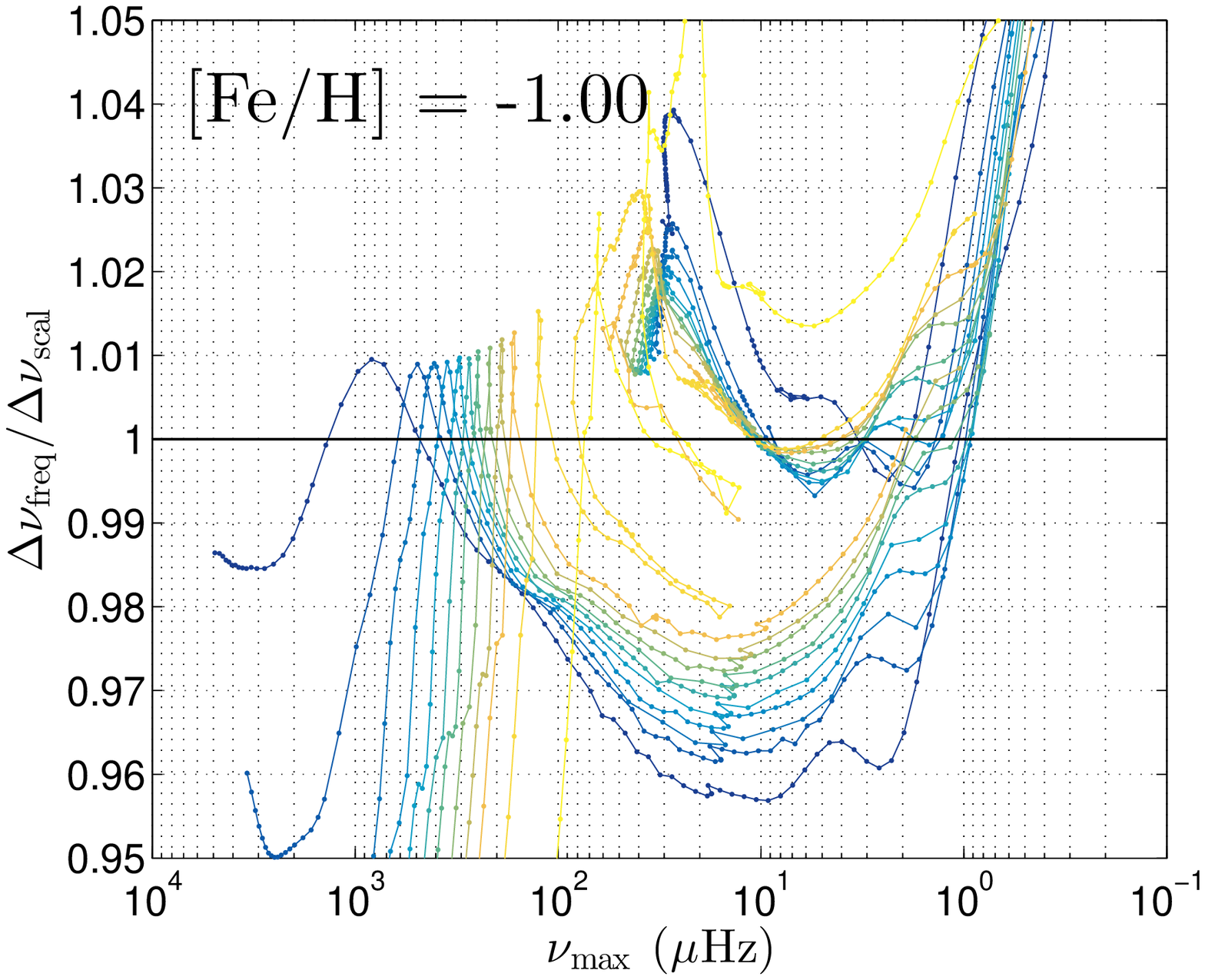}}
    \white{\bf{\\0\\}}
    \resizebox{0.9\textwidth}{!}{\includegraphics{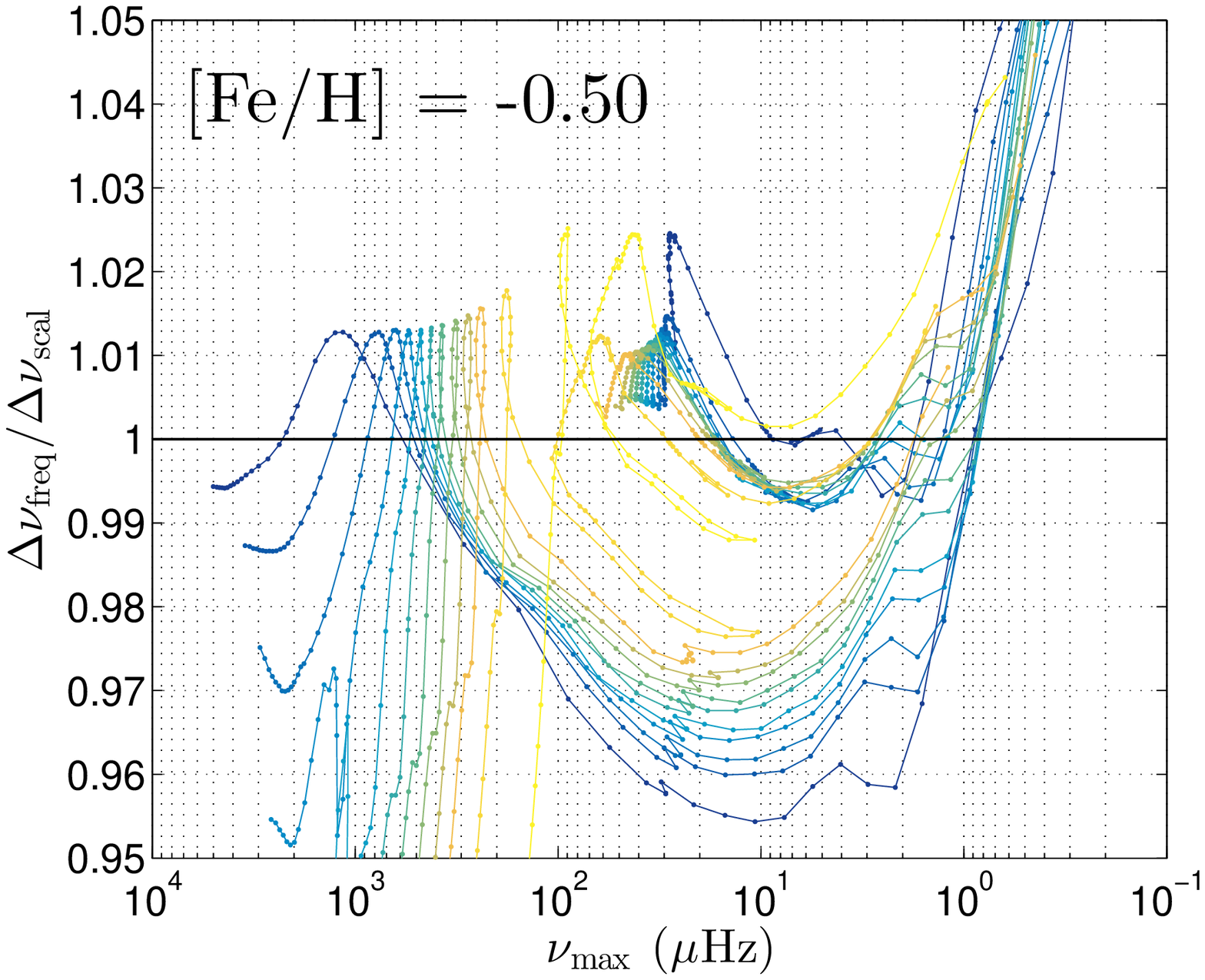}}
    \white{\bf{\\0\\}}
    \resizebox{0.9\textwidth}{!}{\includegraphics{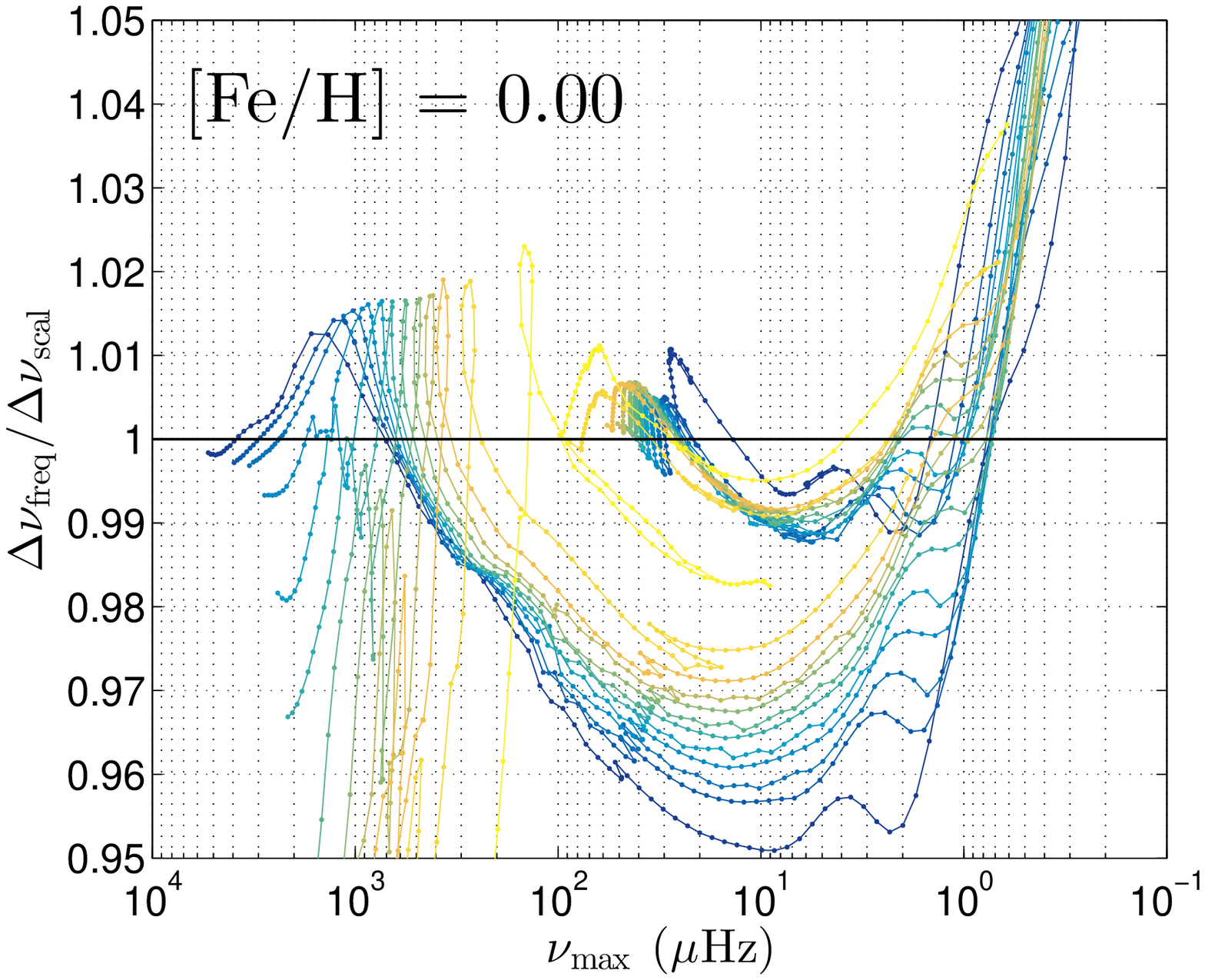}}
    \white{\bf{\\0\\}}
    \resizebox{0.9\textwidth}{!}{\includegraphics{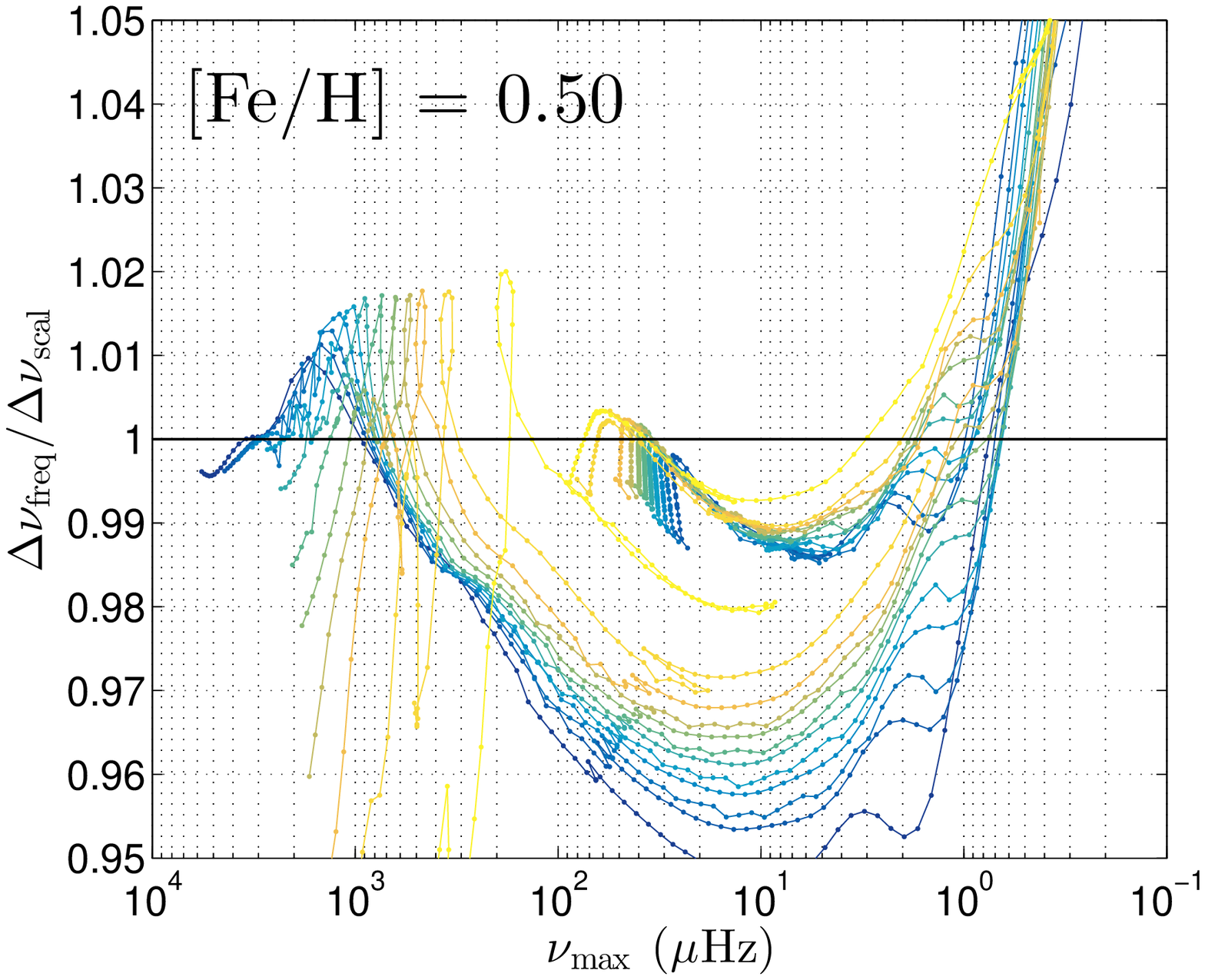}}
    \white{\bf{\\0\\}}
    \end{minipage}
    \caption{Correction of scaling relation \deltanu\ in function of \numax, for a subsection of the grid of tracks presented in Section~\ref{sec:grid}.}
    \label{fig:gridnmaxdnu}
\end{figure*}

Figure \ref{fig:gridnmaxdnu} shows the ratio between the large separation obtained from the scaling relation, $\deltanu_\mathrm{scal.}$, and $\langle\deltanu\rangle$ (calculated as described in Sec. \ref{sec:DetAverageDnu}) as a function of \numax\ for a large number of tracks in our computed grid.  
These panels illustrate the dependence of $\Delta\nu$ corrections on mass, evolutionary state and also chemical composition that affects the mass distribution inside the star. 

As shown in Fig.~\ref{fig:gridnmaxdnu} the deviation of $\Delta\nu$ with respect to the scaling relation tends to low values for stars in the secondary clump. We must keep in mind however, that the masses of stars populating the secondary clump depend on the mixing processes occurred during the previous main-sequence phase, and also on the chemical composition, that is  metallicity and initial mass fraction of He. Therefore, a straightforward parametrization of correction as function of mass and metallicity is not possible. 


\subsection{Period spacing}
\label{sec:deltaP}

It has been shown by \citet{Mosser12a} that is possible to infer the asymptotic period spacing of a star, by fitting a simple pattern on their oscillation spectra. 
This is particularly relevant for those stars that present a rich forest of dipole modes ($l=1$), like, for instance, the red giants.
The asymptotic theory of stellar oscillation tells us that the g-modes are related by an asymptotic relation where their periods are equally spaced by $\deltaP_{l}$.
The relation states that the asymptotic period spacing is proportional to the inverse of the integral of the \brunt\ frequency $N$ inside the trapping cavity: 
\beq
    \deltaP_{l} = \dfrac{2\pi^2}{\sqrt{l(l+1)}}\left(\displaystyle\int^{r_2}_{r_1}\dfrac{N}{r}\mathrm{d} r\right)^{-1},
    \label{eq:DPg}
\eeq
where $r_1$ and $r_2$ are the coordinates in radius of turning points that limit the cavity. 
It is easy to see that its value depends, among other things,  on the size and the position of the internal cavity, fact that will become particularly relevant in the helium-core-burning phase, giving the uncertainties on the core convection \citep{Montalban_etal13,bossini15}.  
On the RGB the period spacing is an excellent tool to set constrains on other stellar quantities, like radius, and luminosity (see for instance \citealt{lagarde16} and \citealt{davies16}).
Moreover the period spacing gives an easy and immediate discrimination between stars in helium-core-burning and in RGB phases, since the former have a \deltaP\ systematically larger of about $\sim 200-300$~s than the latter, while after the early-AGB phase it decreases to similar or smaller values.


\subsection{A quick introduction to grid-based and Bayesian methods}

Having introduced the way \deltanu\ (hereafter, 
to simplify $\deltanu=\langle\deltanu\rangle$) and \deltaP\ are computed in the grids of tracks, let us first remind how they enter in the grid-based and Bayesian methods.

In the so-called direct methods, the asteroseismic quantities are used to provide estimates of stellar parameters and their errors, by directly entering them either in formulas (like the scaling relations of equation~\ref{eq:scaling}) or in 2D diagrams built from grids of stellar models. In grid-based methods with Bayesian inference, this procedure is improved by the weighting of all possible models and by updating the probability with additional information about the data set, described approximately as:
\begin{equation}
p(\mathbf{x}|\mathbf{y})\sim p(\mathbf{y}|\mathbf{x}) p(\mathbf{x}),
\label{eq_ppdf}
\end{equation}
where  $p(\mathbf{x}|\mathbf{y})$ is the posterior probability density function (PDF), $p(\mathbf{y}|\mathbf{x})$ is the likelihood function,
which makes the connection between the measured data $\mathbf{y}$ and the models described as a function of parameters to be derived $\mathbf{x}$, and $p(\mathbf{x})$ is the prior probability function that describes the knowledge about the derived parameters obtained before the measured data. The uncertainties of the measured data are usually described as a normal distribution, therefore the likelihood function is written as
\begin{equation}
p(\mathbf{y^\prime}|\mathbf{x}) = \prod_i \frac{1}{\sqrt{2\pi}\sigma_{y_i}}  \times \exp{\left(\frac{-(y_i^\prime-y_i)^2}{2\sigma_{y_i}^2} \right)},
\label{eg:likelihood}
\end{equation}
where ${y_i^\prime}$ and $\sigma_{y_i^\prime}$ are the mean and standard deviation, for each of the $i$ quantities considered in the data set.

In order to obtain the stellar quantity $x_i$, the posterior PDF is then integrated over all parameters, except $x_i$, resulting a PDF for this parameter. For each PDF, a central tendency (mean, mode, or median) is calculated with their credible intervals. Therefore this method requires not only trusting the stellar evolutionary models but also adopting a minimum set of reasonable priors (in stellar age, mass, etc.). In addition to avoid the scaling relations, the method requires that the asteroseismic quantities are tabulated along a set of stellar models, covering the complete relevant interval of masses, ages, and metallicities.


\subsection{Interpolating the \texorpdfstring{\deltanu}{Dnu} deviations to make isochrones}

\begin{figure}
    \resizebox{0.85\hsize}{!}{\includegraphics{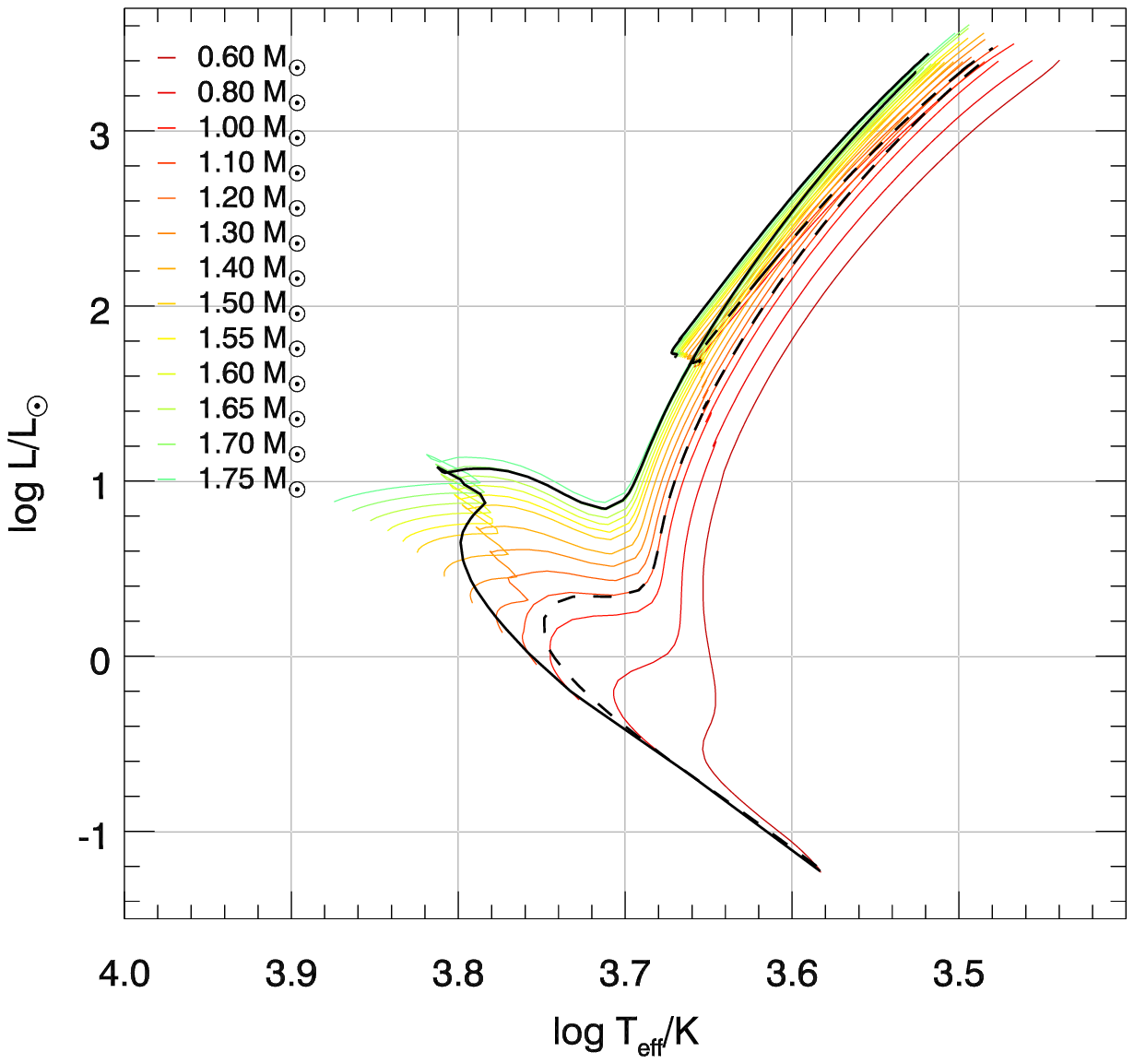}} 
    \resizebox{0.85\hsize}{!}{\includegraphics{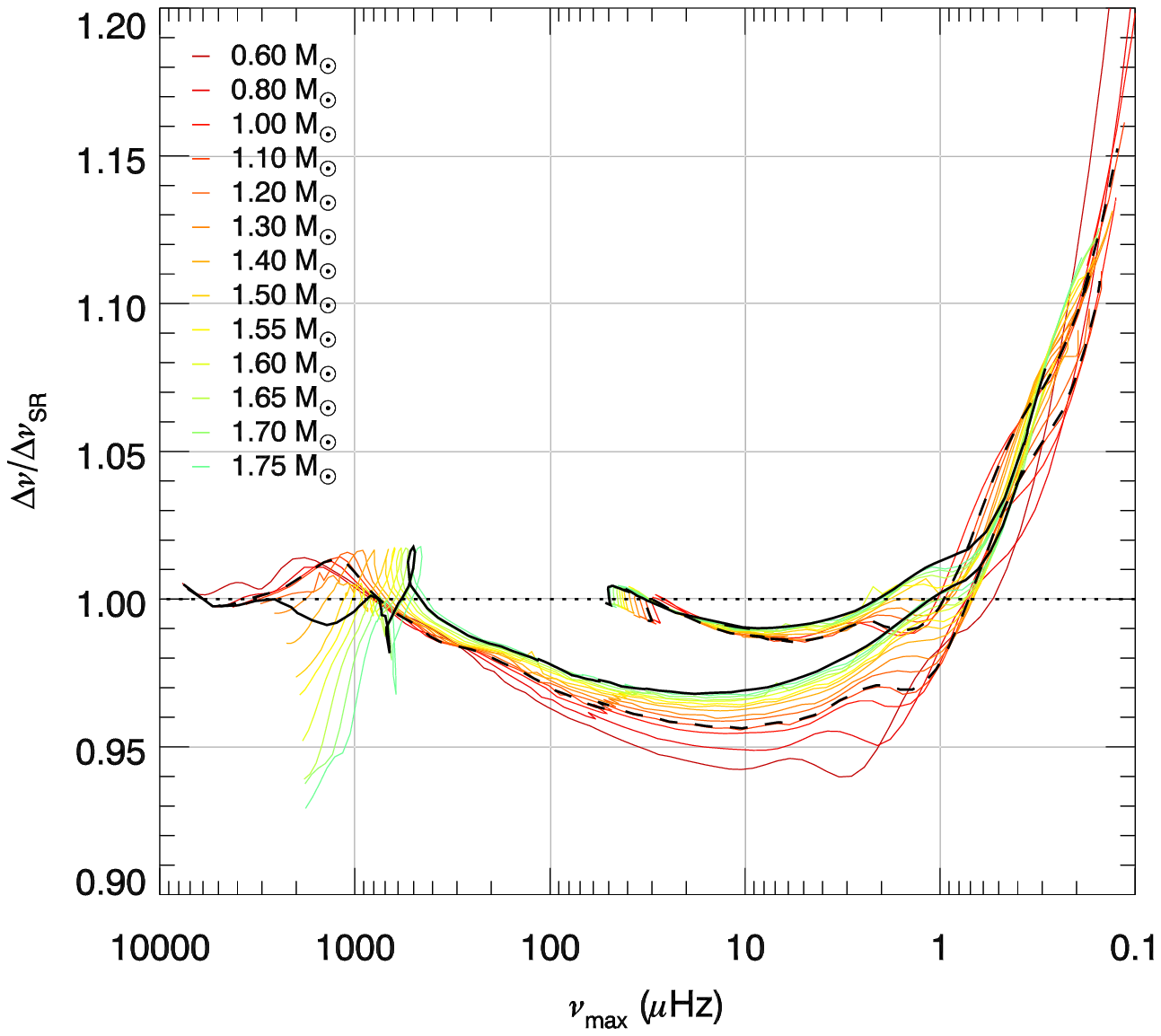}}
    \resizebox{0.85\hsize}{!}{\includegraphics{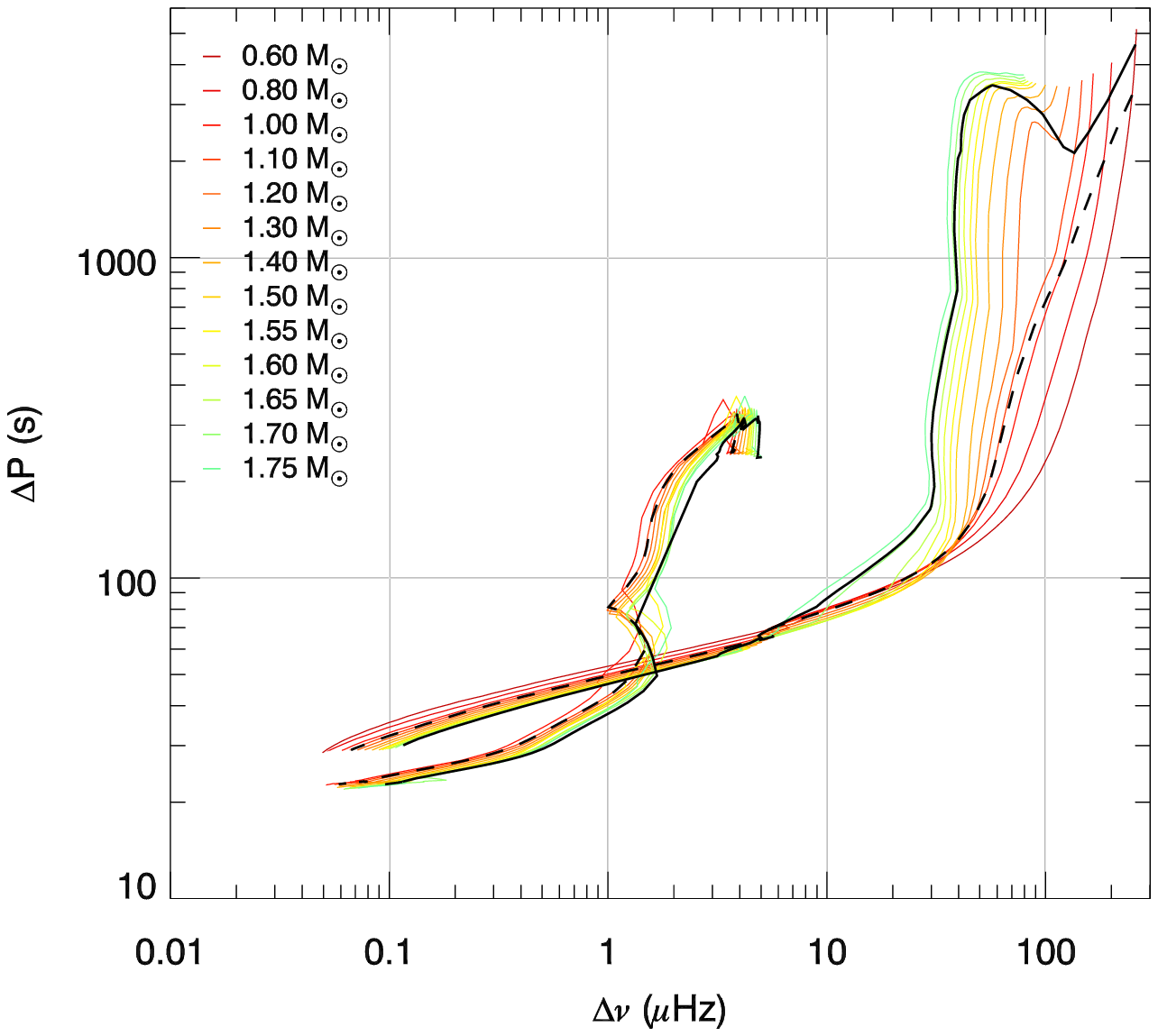}}
    \caption{MESA evolutionary tracks color coded according to mass in the HR (top panel), $\deltanu/\deltanu_\text{SR}$ {\it versus} \numax (middle), and \deltaP\ {\it versus} \deltanu\ (bottom) diagrams. The solid and the dashed black lines are examples of interpolated isochrones of 2 and 10~Gyr, respectively.}
    \label{fig_tracks_isoc}
\end{figure}

The \deltanu\ computed along the tracks appropriately sample stars in the most relevant evolutionary stages, and over the interval of mass and metallicity to be considered in this work. However, in order to be useful in Bayesian codes, a further step is necessary: such calculations need to be interpolated for {\em any} intermediate value of evolutionary stage, mass, and metallicity. This would allow us to derive detailed isochrones, that can enter easily in any estimation code which involves age as a parameter. Needless to say, such isochrones may find many other applications.

The computational framework to perform such interpolations is already present in our isochrone-making routines, which are described elsewhere \citep[see][]{marigo16}. In short, the following steps are performed: our code reads the evolutionary tracks of all available initial masses and metallicities; these tracks contain age ($\tau$), luminosity ($L$), \Teff, \deltanu, and \deltaP\ from the ZAMS until TP-AGB. These quantities are interpolated between the tracks, for any intermediate value of initial mass and metallicity, by performing linear interpolations between pairs of ``equivalent evolutionary points'', i.e.,\ points in neighbouring tracks which share similar evolutionary properties. An isochrone is then built by simply selecting a set of interpolated points for the same age and metallicity. In the case of \deltanu, the interpolation is done in the quantity $\deltanu/\deltanu_\text{SR}$, where $\deltanu_\text{SR}$ is the value defined by the scaling relation in equation~\ref{eq:scaling}. In fact, $\deltanu/\deltanu_\text{SR}$ varies along the tracks in a much smoother way, and has a much more limited range of values than the $\deltanu$ itself; therefore the multiple interpolations of its value among the tracks also produce well-behaved results. Of course, in the end the interpolated values of $\deltanu/\deltanu_\text{SR}$ are converted into $\deltanu$, for every point in the generated isochrones.

Figure \ref{fig_tracks_isoc} shows a set of evolutionary tracks until the TP-AGB phase in the range $[0.60,1.75]$~\Msun\ for $\feh=0.25$ ($Z=0.03123$) and interpolated isochrones of 2 and 10~Gyr both in the Hertzsprung--Russell (HR), the ratio $\deltanu/\deltanu_\text{SR}$ {\it versus} \numax, and the \deltaP\ {\it versus} \deltanu\ diagrams. The middle panel shows the deviation of the scaling \deltanu\ of few percents mainly over the RGB and early-AGB phases. Deviations at the stages of main sequence and core helium burning are generally smaller than one per cent.

The Fig.~\ref{fig_tracks_isoc} also shows that our interpolation scheme works very well, with the derived isochrones reproducing the behaviour expected from the evolutionary tracks.

No similar procedure was necessary for the interpolation in \deltaP, since it does not follow any simple scaling relation, and it varies much more smoothly and covering a smaller total range than \deltanu. The interpolations of \deltaP\ are simply linear ones using parameters such as mass, age (along the tracks), and initial metallicity as the independent parameters.


\section{Applications}
\label{sec:applications}

We derived the stellar properties using the Bayesian tool PARAM \citep{dasilva06, rodrigues14}. From the measured data -- \Teff, \mh, \deltanu, and \numax -- the code computes PDFs for the stellar parameters: $M$, $R$, $\log{g}$, mean density, absolute magnitudes in several passbands, and as a second step, it combines apparent and absolute magnitudes to derive extinctions $A_V$ in the $V$ band and distances $d$. The code uses a flat prior for metallicity and age, while for the mass, the \citet{chabrier01} initial mass function was adopted with a correction for small amount of mass lost near the tip of the RGB computed from \citet{reimers75} law with efficiency parameter $\eta=0.2$ \citep[cf.][]{miglio12_2}. The code also has a prior on evolutionary stage that, when applied, separates the isochrones into 3 groups: ‘core-He burners’ (RC), ‘non-core He burners’ (RGB/AGB), and only RGB (till the tip of the RGB). The statistical method and some applications are described in details in \citet{rodrigues14}. 

We expanded the code to read the additional seismic information of the MESA models described in Section~\ref{sec:models}. We implemented new variables to be taken into account in the likelihood function (equation~\ref{eg:likelihood}), such as \deltanu\ from the model frequencies, \deltaP, $\log{g}$, and luminosity. Hence the entire set of measured data is
\begin{equation}
\mathbf{y}=(\mh,\Teff,\deltanu,\numax,\deltaP,\log{g},L), \nonumber
\end{equation}
where \deltanu\ can still be computed using the standard scaling relation (hereafter \deltanu(SR)).
Therefore PARAM is now able to compute stellar properties using several different input configurations, i.e., the code can be set to use different combinations of measured data. Some interesting cases are, together with \Teff\ and \mh,
\begin{itemize}
\item \deltanu\ and \numax\ from scaling relation (equation~\ref{eq:scaling});
\item \deltanu\ from model frequencies and \numax\ from scaling relation; 
\item \deltanu\ (either from model frequencies or scaling relation), together with some other asteroseismic parameter, such as \deltaP;
\item $\log{g}$;
\item any of the previous options together with the addition of a constraint on the stellar luminosity.
\end{itemize}
The first two cases constitute the main improvement we consider in this paper, which is already subject of significant attention in the literature \citep[see e.g.][]{sharma16, guggenberger16}. The third case is particularly important given the fact that the \numax\ scaling relation is basically empirical and may still reveal small offsets in the future.  
Finally, the fourth and fifth cases are aimed at exploring the effect of lacking of seismic information, when only spectroscopic data is available for a given star; and adding independent information in the method, like e.g.\ the known distance of a cluster, or of upcoming Gaia parallaxes, respectively.


\subsection{Tests with artificial data}
\label{sec:artificial}

To test the precision that we could reach with a typical set of observational constraints available for {\it Kepler} stars, we have chosen 6 models from our grid of models and considered various combinations of seismic, astrometric, and spectroscopic constraints (see Table \ref{tab:artificial}).

The seismic constraints taken from the artificial data are \deltanu, \numax, and \deltaP. The latter is used by taking its asymptotic value as additional constraint in equation~\ref{eg:likelihood}, and not as only a discriminant for the evolutionary phase as done in previous works \citep[e.g.][]{rodrigues14}. Uncertainties on \deltanu\ and \numax\ were taken from \citet{handberg16} and on \deltaP\ from \citet{Vrard2016}. We adopted 0.2~dex as uncertainties on $\log{g}$ based on average values coming from spectroscopy. For luminosity, we adopted uncertainties of the order of 3 per cent based on Gaia parallaxes,  where a significant fraction of the uncertainty comes from bolometric corrections \citep{reese16}.

\begin{figure*}
    \begin{minipage}{\columnwidth}
    \resizebox{\hsize}{!}{\includegraphics{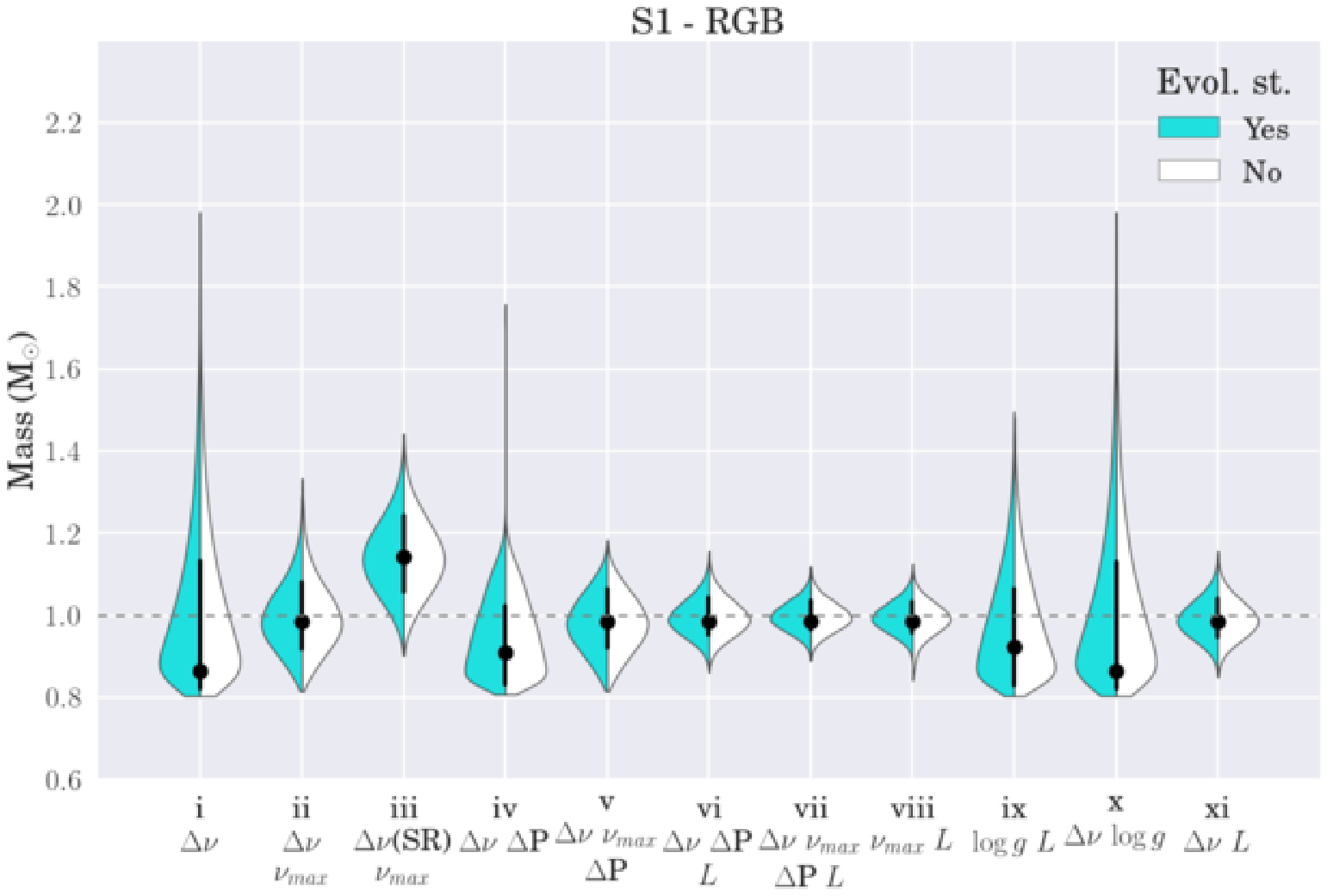}} \\
    \resizebox{\hsize}{!}{\includegraphics{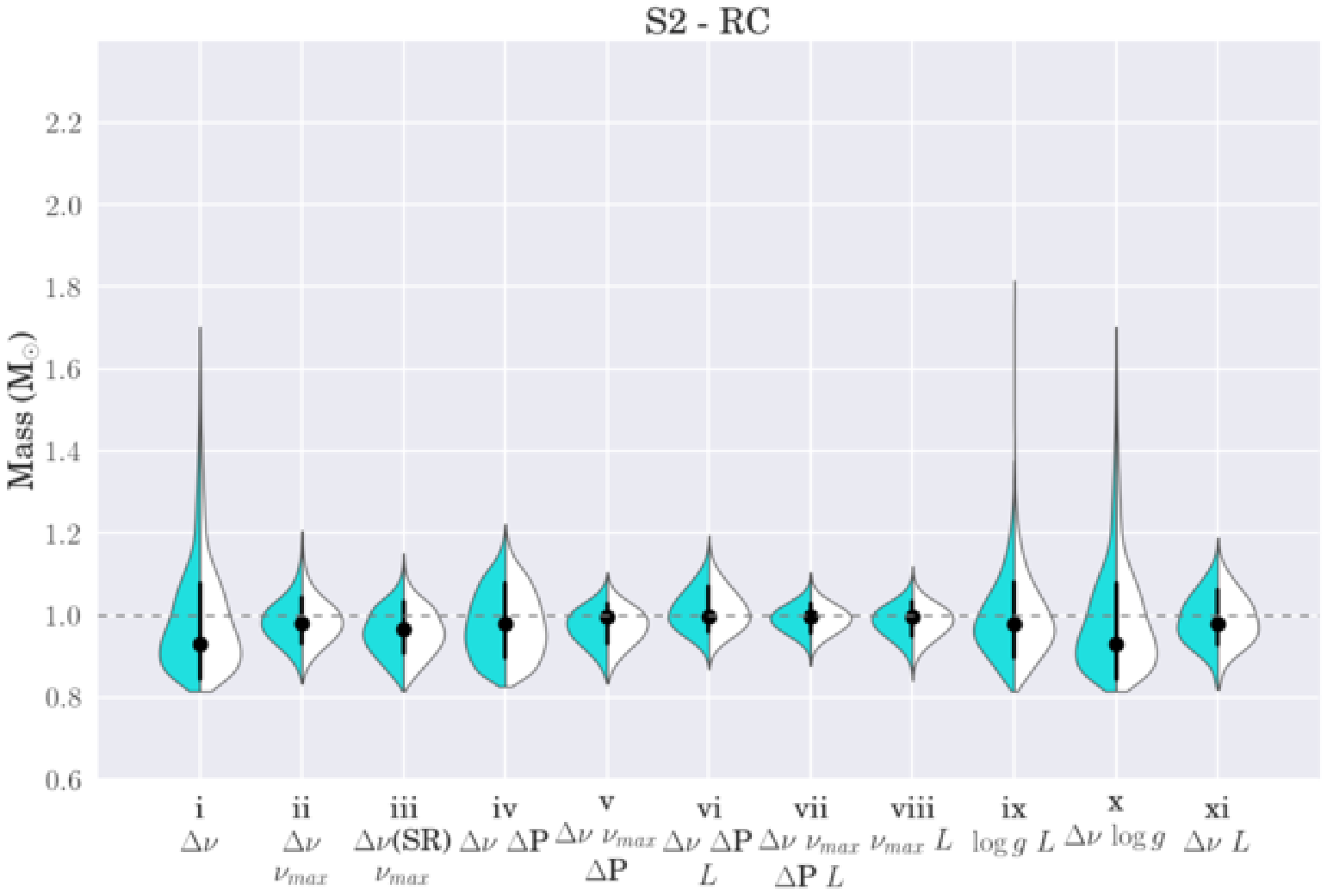}} \\
    \resizebox{\hsize}{!}{\includegraphics{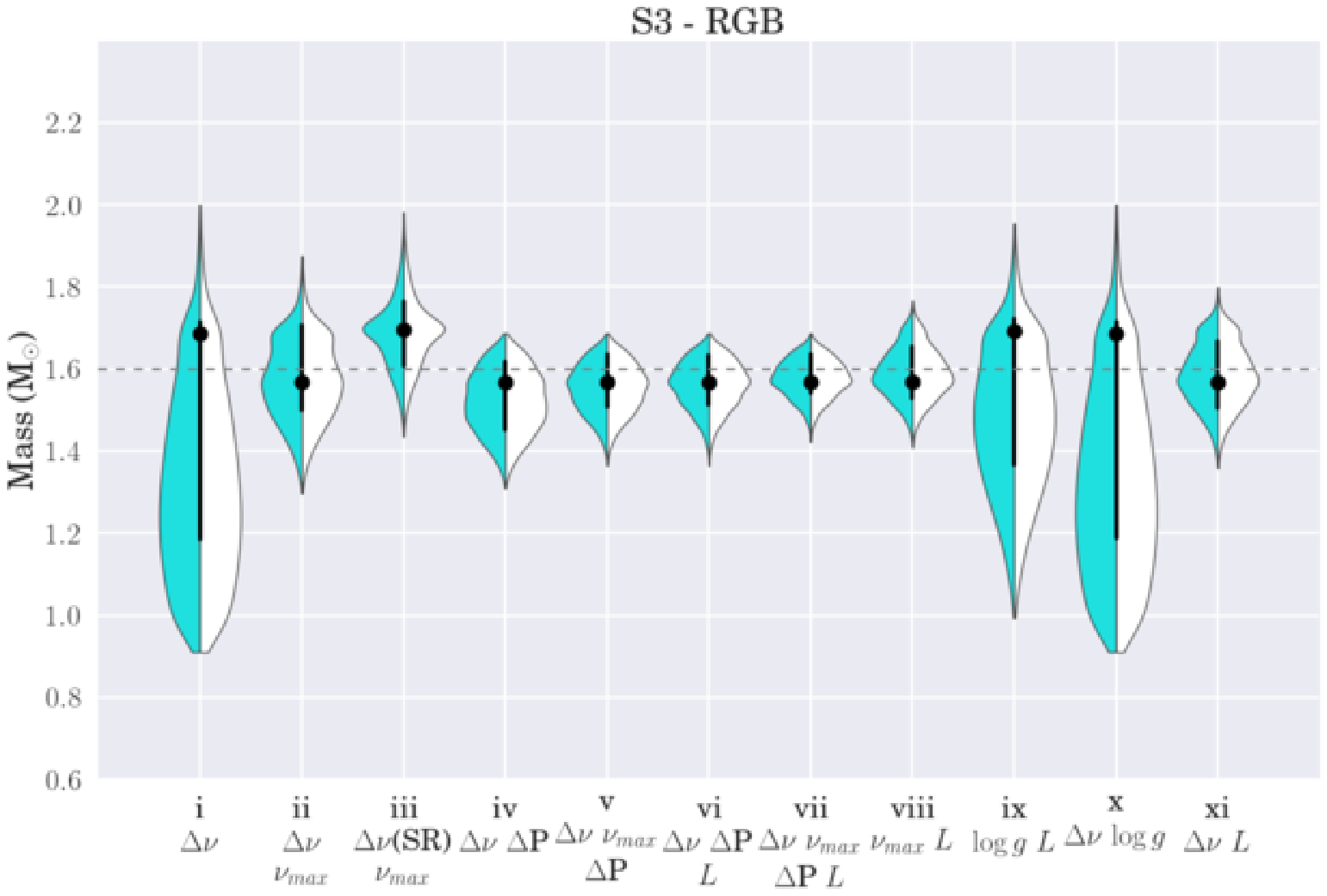}} 
    \end{minipage}
    \begin{minipage}{\columnwidth}
    \resizebox{\hsize}{!}{\includegraphics{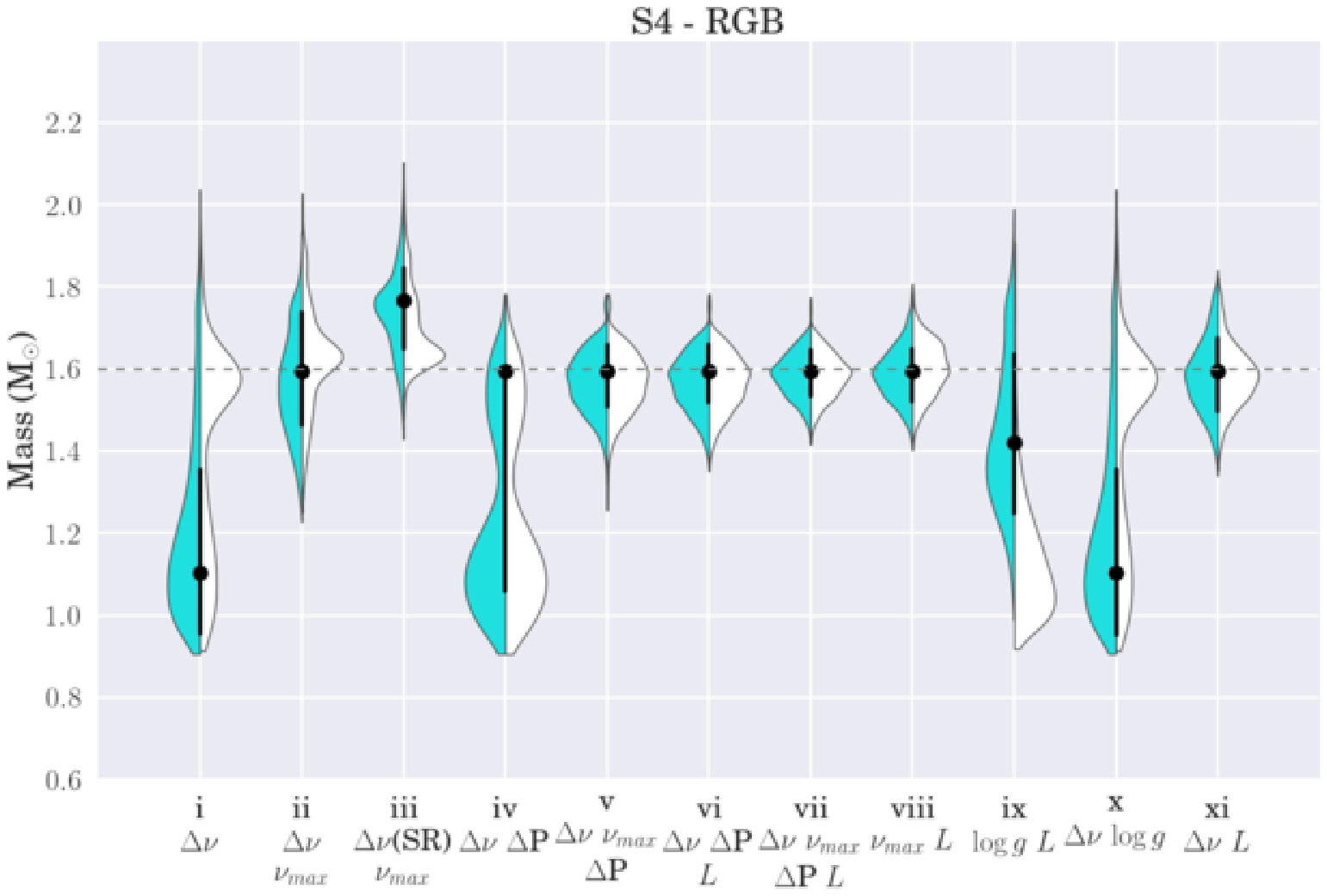}} \\
    \resizebox{\hsize}{!}{\includegraphics{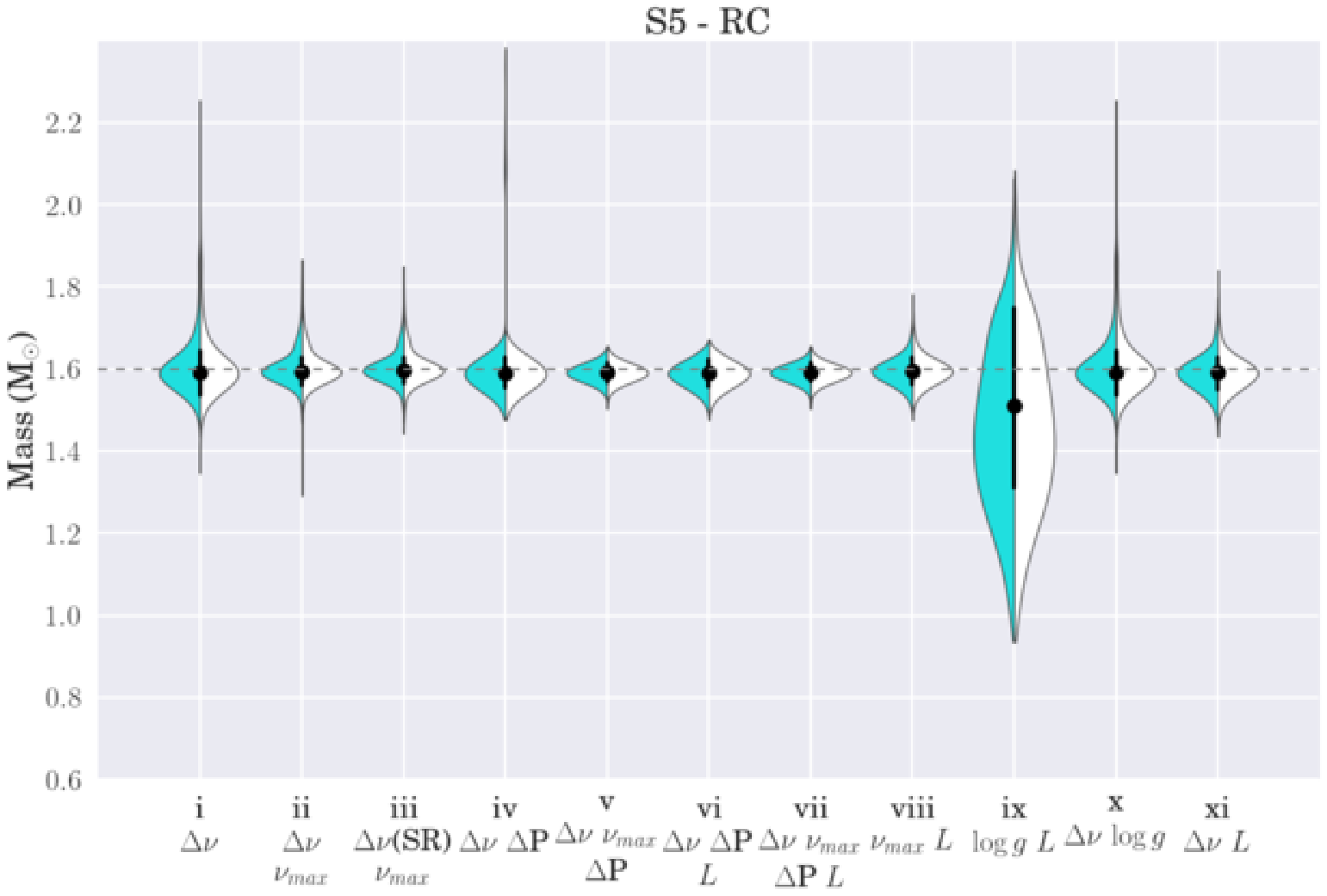}} \\
    \resizebox{\hsize}{!}{\includegraphics{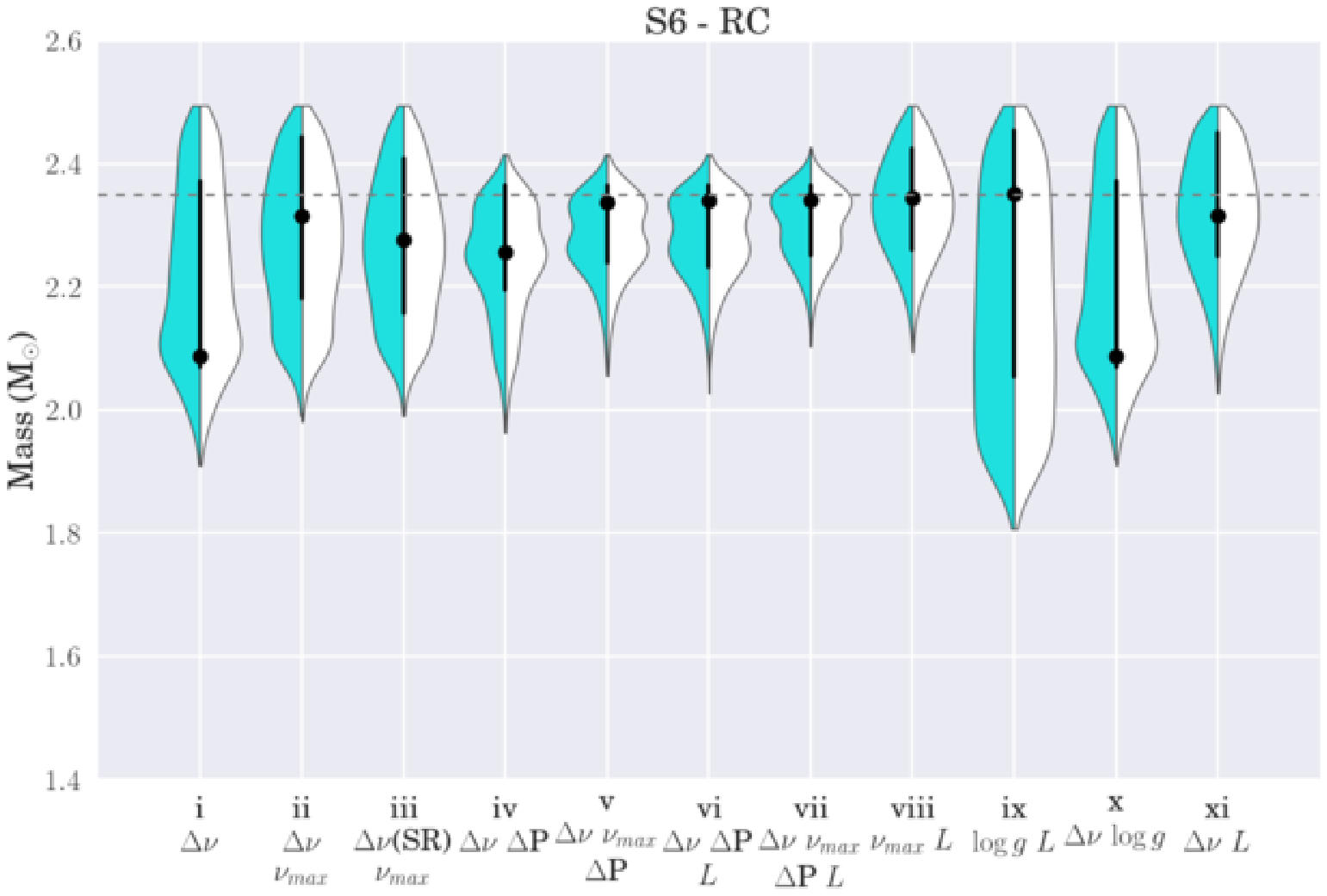}}
    \end{minipage}
\caption{PDFs of mass for the 6 artificial stars presented in Table~\ref{tab:artificial} using violin plots. Each panel shows the results of one star, named in the top together with its evolutionary stage. The $x$ axis indicates each combination of input parameters for PARAM code as described in Section~\ref{sec:artificial}. The left side of the violin (cyan color) represents the resulting PDF when prior on evolutionary stage is applied, while in the right side (white color) the prior is not being used. The black dots and error bars represent the mode and its 68 per cent credible intervals of the PDF with prior on evolutionary stage (cyan distributions). The dashed line indicates the mass of the artificial stars.}
\label{fig:pdfmass}
\end{figure*}

\begin{figure*}
    \begin{minipage}{\columnwidth}
    \resizebox{\hsize}{!}{\includegraphics{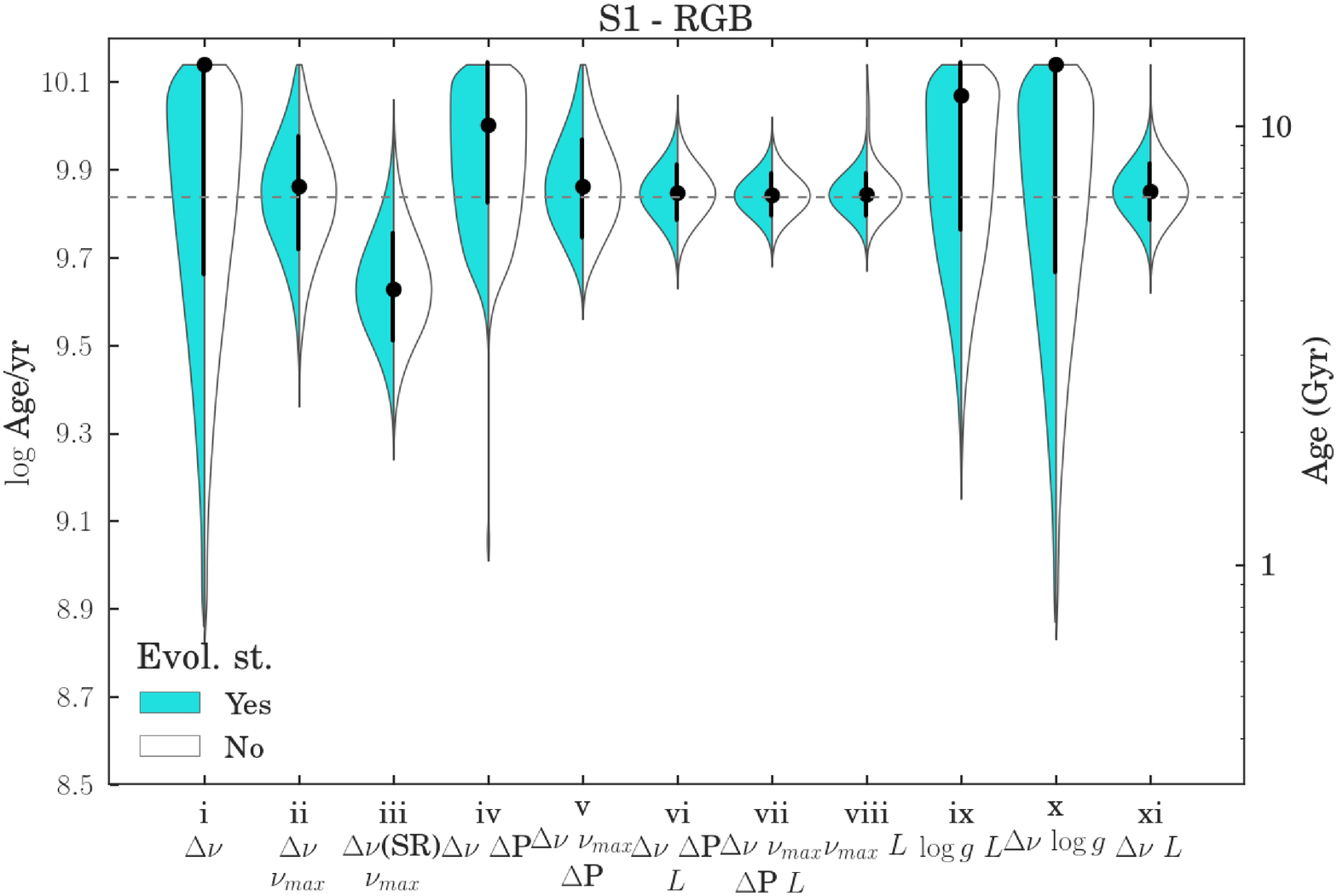}} \\
    \resizebox{\hsize}{!}{\includegraphics{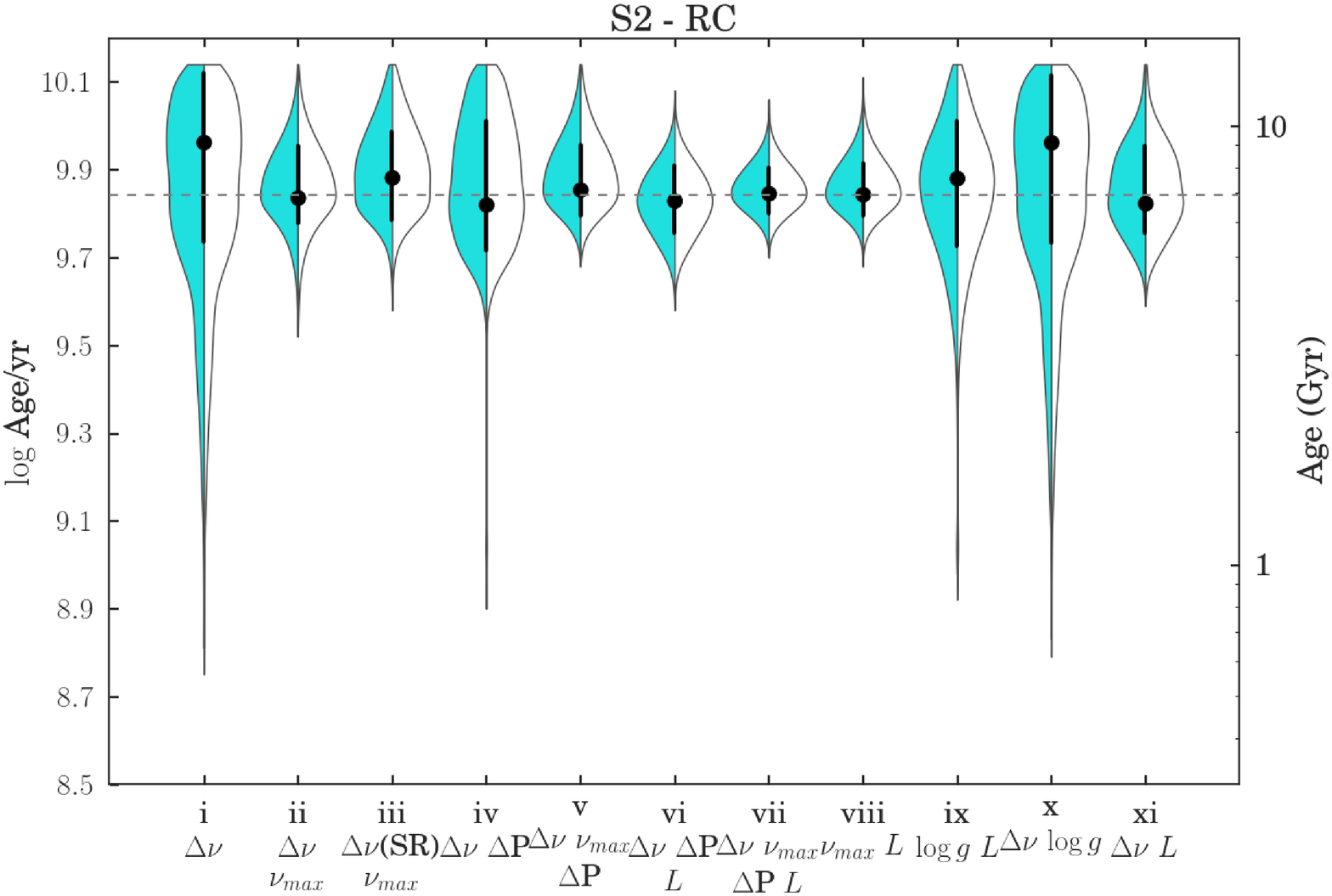}} \\
    \resizebox{\hsize}{!}{\includegraphics{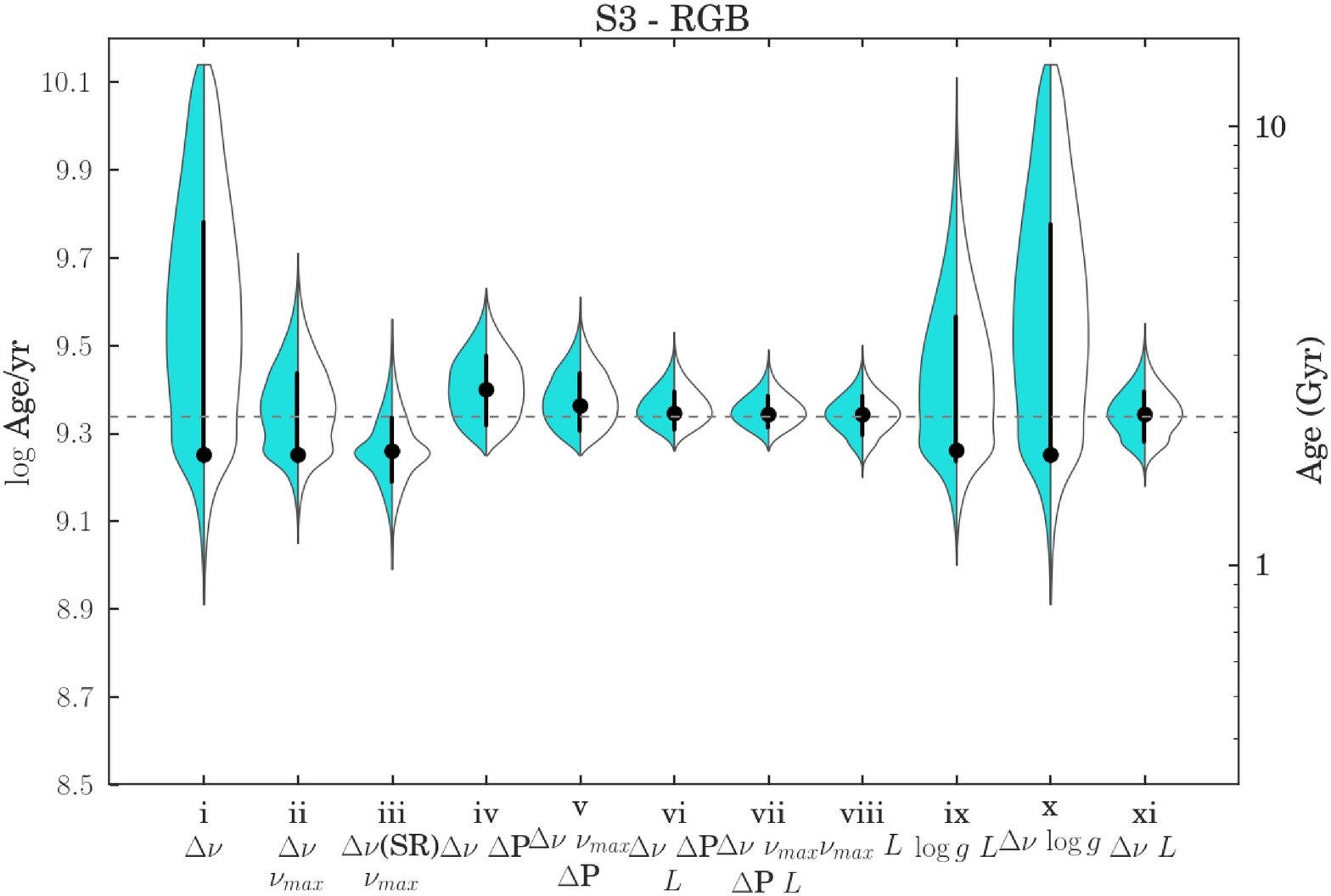}} 
    \end{minipage}
    \begin{minipage}{\columnwidth}
    \resizebox{\hsize}{!}{\includegraphics{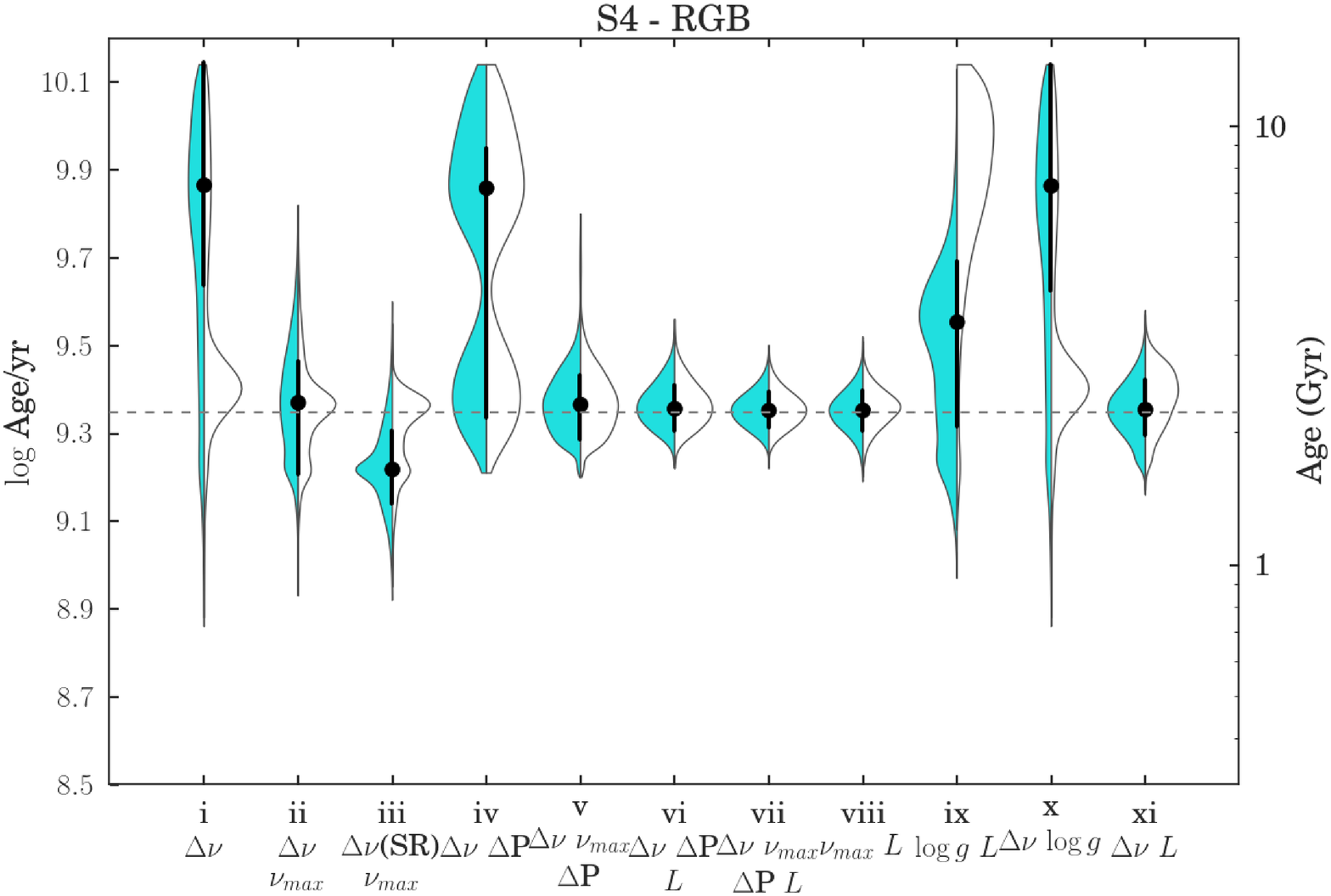}} \\
    \resizebox{\hsize}{!}{\includegraphics{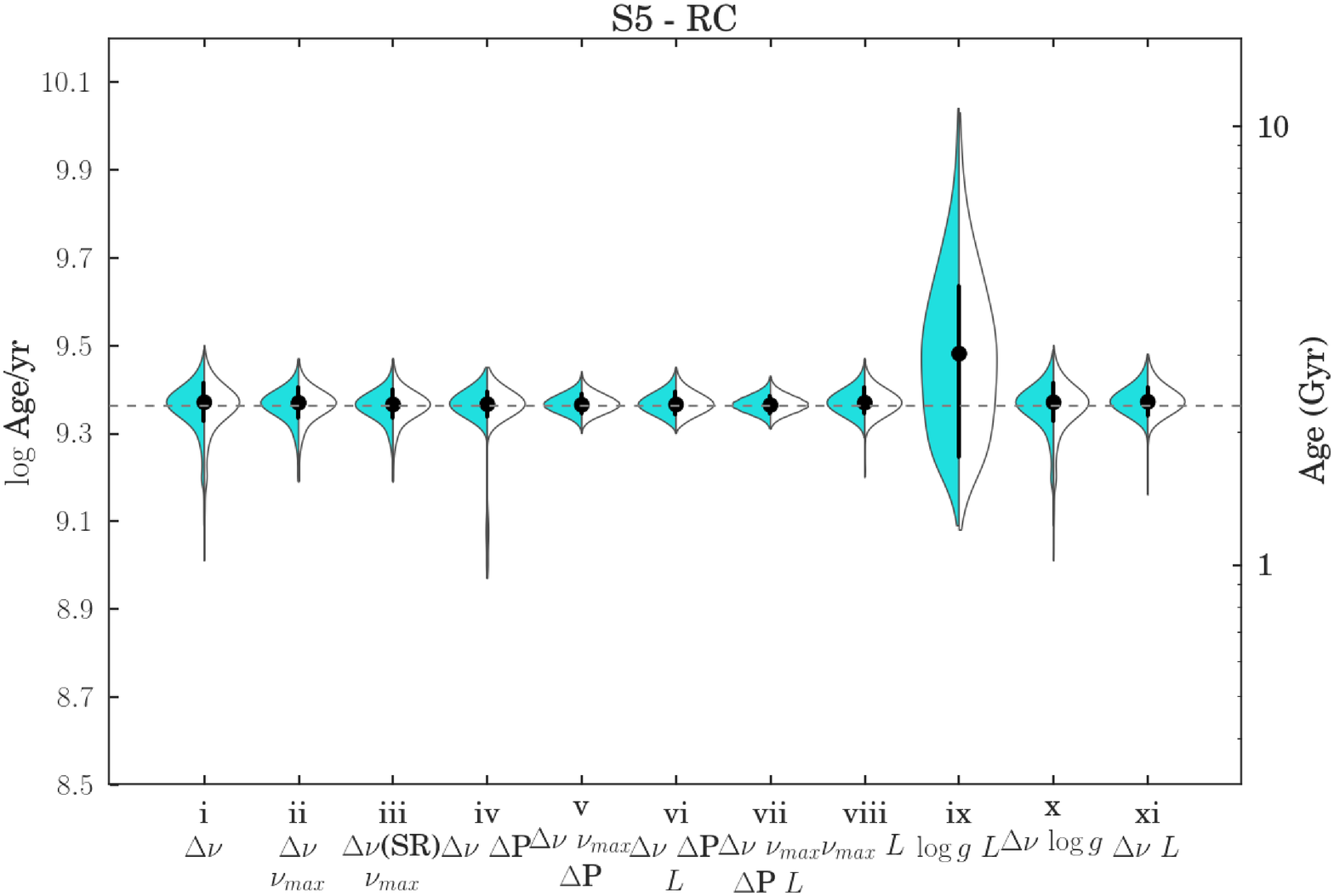}} \\
    \resizebox{\hsize}{!}{\includegraphics{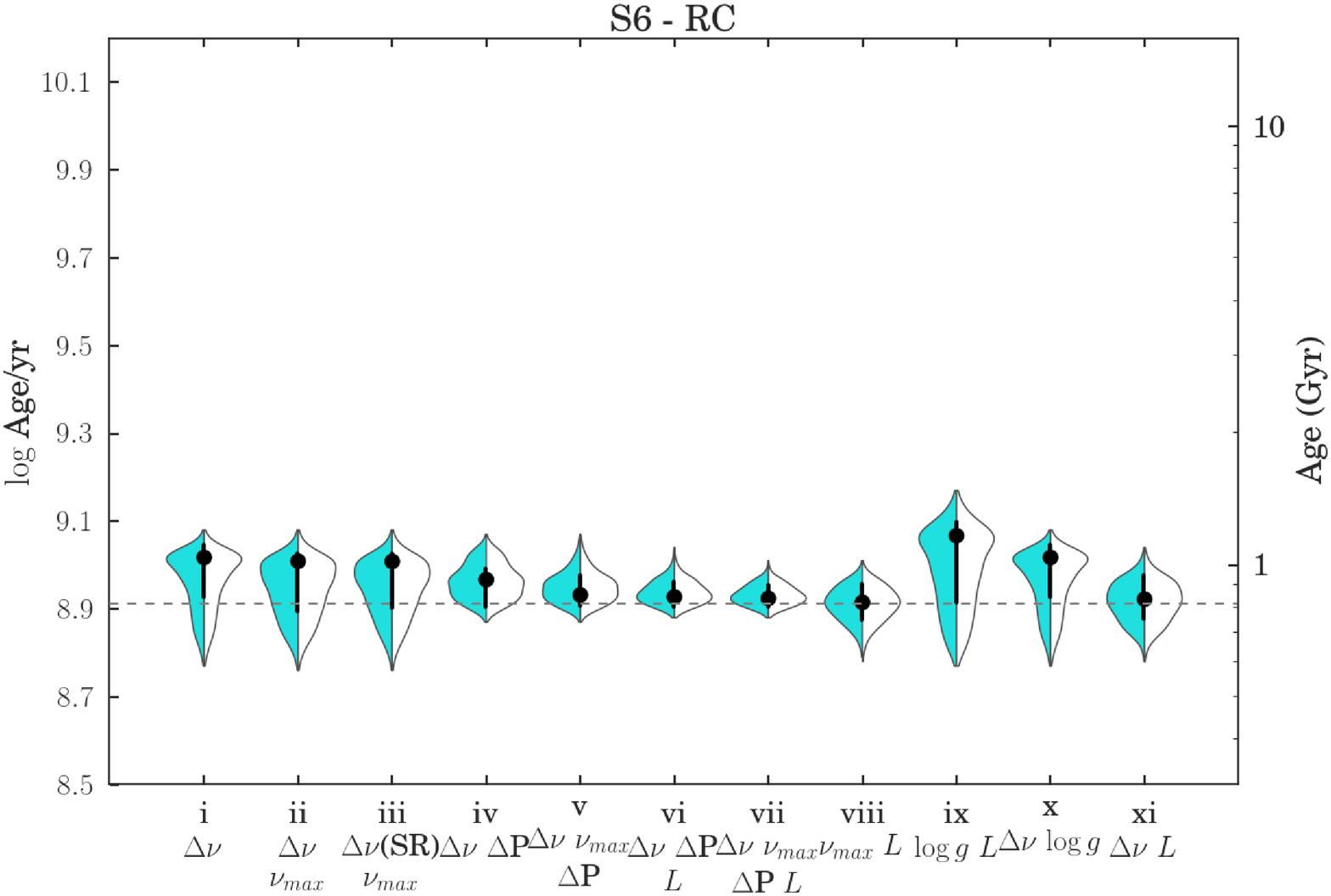}}
    \end{minipage}
\caption{The same as figure~\ref{fig:pdfmass} but for logarithm of the ages. The right y-axis gives the age in Gyr. The dashed line indicates the age of the artificial stars.}
\label{fig:pdfage}
\end{figure*}

We derived stellar properties using 11 different combinations as input to PARAM, in all cases using \Teff\ and \feh, explained as following:
\begin{enumerate}[i]
\item \deltanu\ -- only \deltanu\ from model frequencies; \label{item:first}
\item \deltanu\ and \numax\ -- to compare with the previous item in order to test if we can eliminate the usage of \numax; \label{item:second}
\item \deltanu(SR) and \numax\ -- traditional scaling relations, to compare with the previous item and correct the offset introduced by using \deltanu\ scaling; \label{item:third}
\item \deltanu\ and \deltaP\ --  in order to test if we can eliminate the usage of \numax\ and improve precision using the period spacing not only as prior, but as a measured data; \label{item:fourth}
\item \deltanu, \numax, and \deltaP\ -- using all the asteroseismic data available; \label{item:fifth}
\item \deltanu, \deltaP, and $L$ -- in order to test if we can eliminate the usage of \numax, when luminosity is available (from the photometry plus parallaxes); \label{item:sixth}
\item \deltanu, \numax, \deltaP, and $L$ -- using all the asteroseismic data available and luminosity, simulating future data available for stars with seismic data observed by Gaia; \label{item:seventh}
\item \numax\ and $L$ -- in the case when it may not always be possible to derive \deltanu\ from lightcurves, simulating possible data from K2 and Gaia surveys; \label{item:eighth}
\item $\log{g}$ and $L$ --  in the case when only spectroscopic data are available (in addition to $L$); \label{item:ninth}
\item \deltanu\ and $\log{g}$ --  again in order to test if we can eliminate the usage of \numax, replacing it by the spectroscopic $\log g$; \label{item:tenth}
\item \deltanu\ and $L$ --  again in order to test if we can eliminate the usage of \numax, when luminosity is available. \label{item:eleventh}
\end{enumerate}
In all cases, the prior on evolutionary stage was also tested. The resulting mass and age PDFs for each artificial star are presented using violin plots\footnote{Violin plots are similar to box plots, but showing the smoothed probability density function.} in Figure~\ref{fig:pdfmass} and \ref{fig:pdfage}, respectively. The $x$ axis indicates each combination of input parameters, as discussed before; the left side of the violin (cyan color) represents the resulting PDF when prior on evolutionary stage is applied, while in the right side (white color) the prior is not being used. The black dots and error bars represent the mode and its 68 per cent credible intervals of the PDF with prior on evolutionary stage (cyan distributions).

\begin{figure*}
    \begin{minipage}{\columnwidth}
    \resizebox{\hsize}{!}{\includegraphics{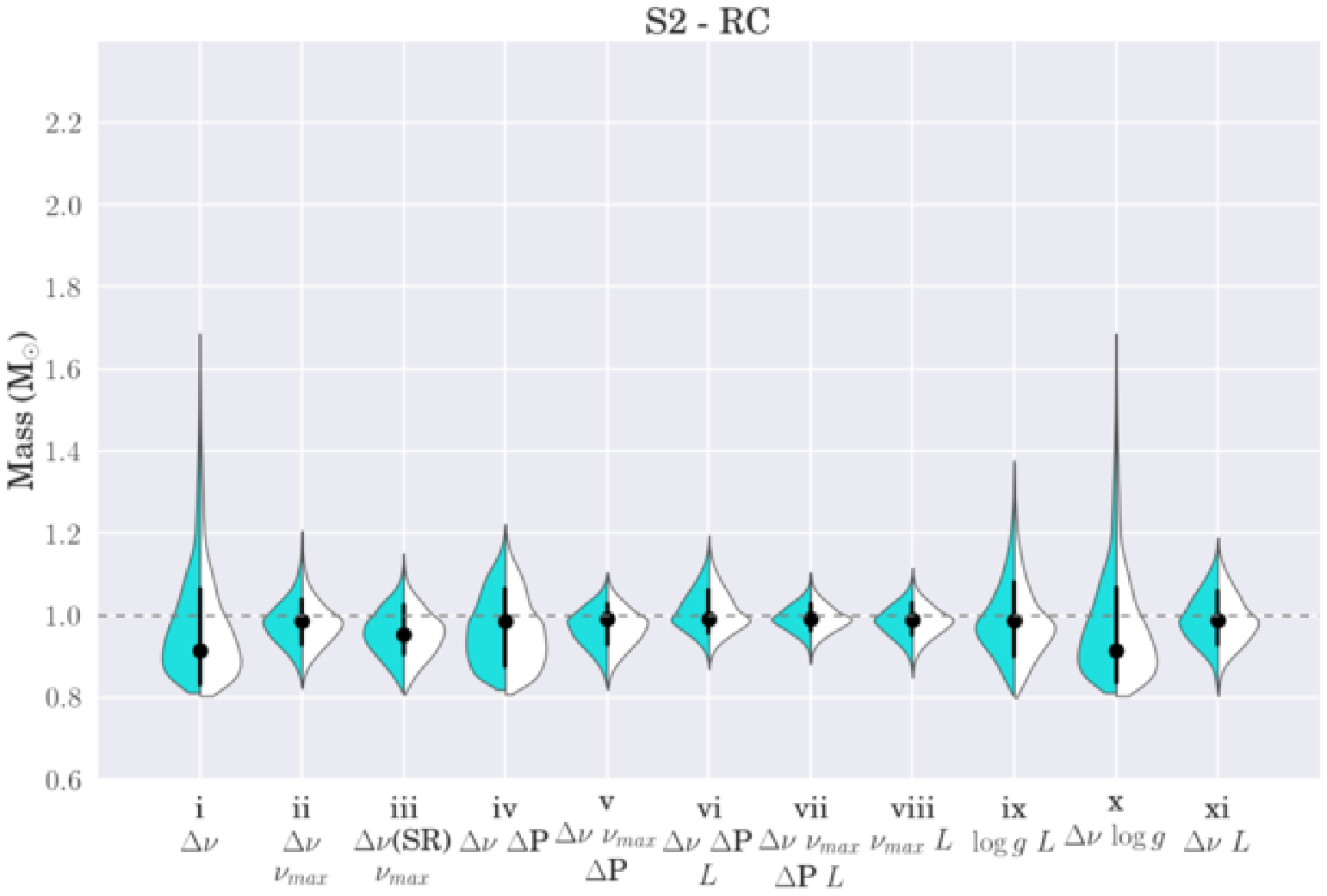}} \\
    \resizebox{\hsize}{!}{\includegraphics{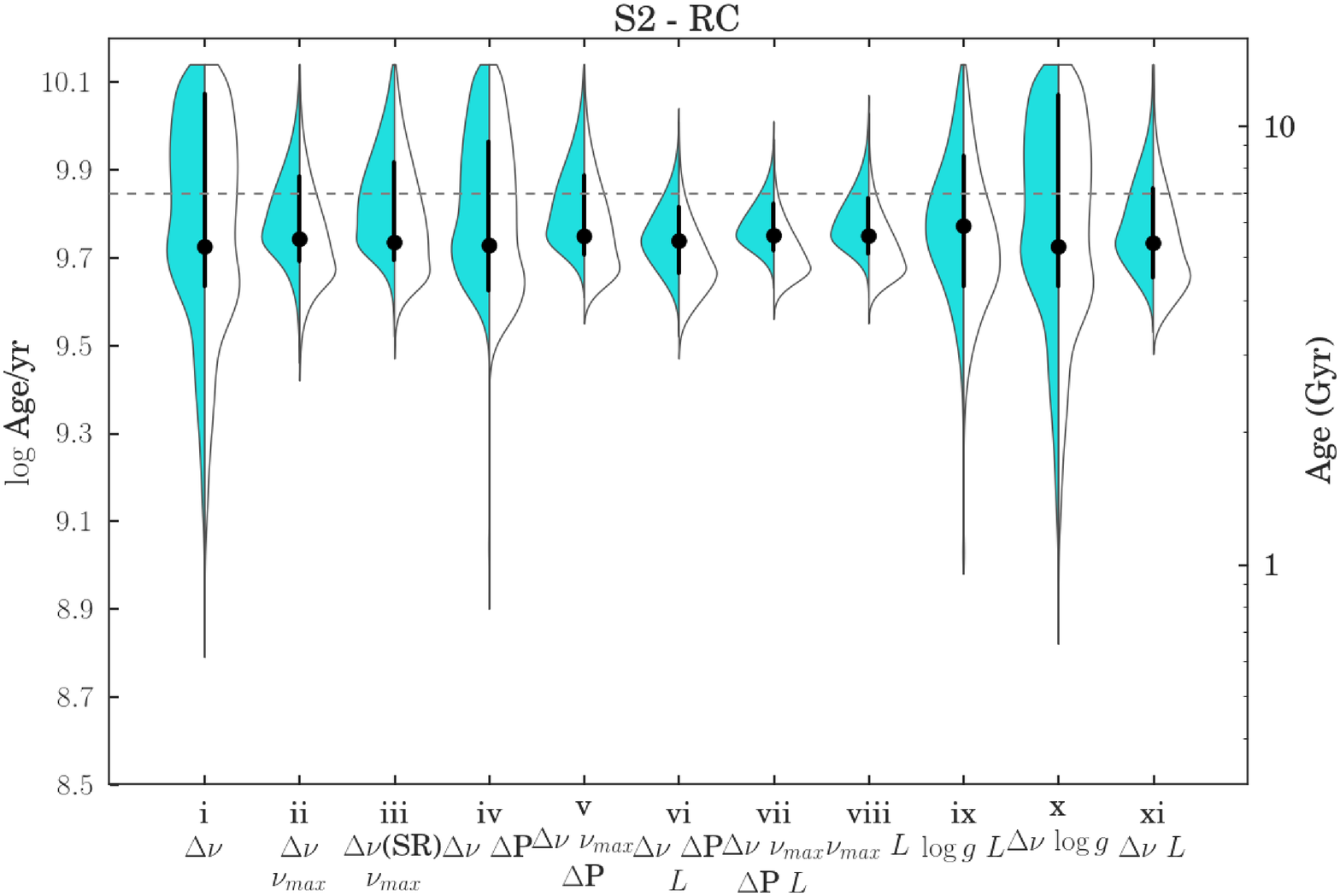}} \\
    \end{minipage}
    \begin{minipage}{\columnwidth}
    \resizebox{\hsize}{!}{\includegraphics{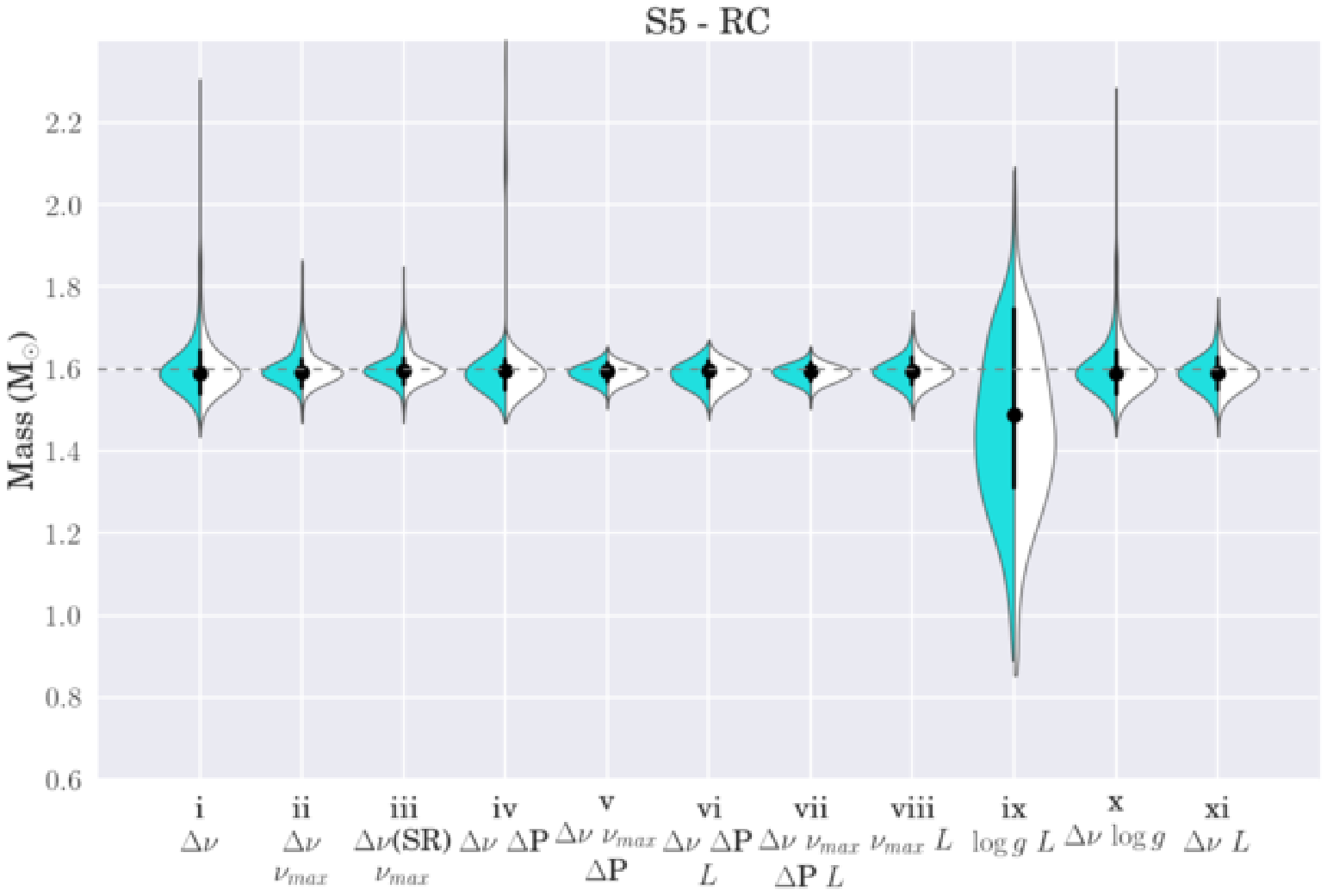}} \\
    \resizebox{\hsize}{!}{\includegraphics{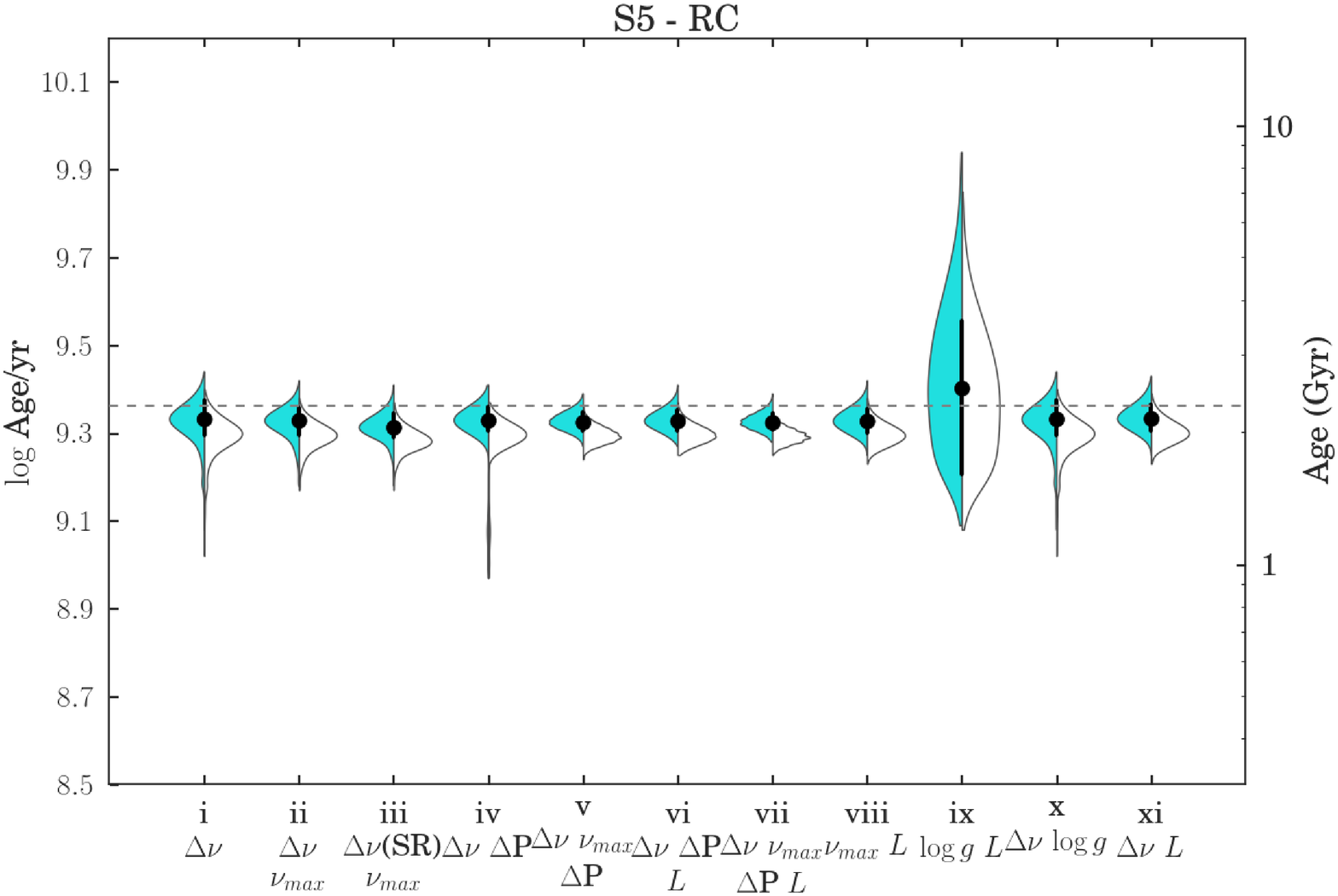}} \\
    \end{minipage}
\caption{PDFs of mass (top panels) and ages (bottom panels) for the artificial RC stars S2 and S5 presented in Table~\ref{tab:artificial} using violin plots. The left side of the violin (cyan color) represents the resulting PDF with the efficiency parameter on mass loss $\eta=0.2$, while in the right side (white color) $\eta=0.4$. The black dots and error bars represent the mode and its 68 per cent credible intervals of the PDF with $\eta=0.2$ (cyan distributions). The dashed line indicates the mass and the ages of the artificial stars.}
\label{fig:pdf_m_age_eta}
\end{figure*}

\begin{table*}
\scriptsize
\centering
\caption{Set of artificial data considered in Section~\ref{sec:artificial}.}
\setlength{\tabcolsep}{4pt}
    \begin{tabular}{ ccc|ccccccccc }
    \hline     Label  &    $M$/M$_\odot$  & $\log$Age/yr & \Teff\ (K) & \feh & $\log{g}$ &$L$/L$_\odot$ & \numax\, ($\mu$Hz)& \deltanu\, ($\mu$Hz)& \deltaP\, (s) & Ev. State \\
    \hline
    S1  & 1.00  & 9.8379  & 4813$\pm$70  &  -0.75$\pm$0.1  & 2.38$\pm$0.20 & 54.77$\pm$1.64 & 30.26$\pm$0.58 & 3.76$\pm$0.05 & 61.40$\pm$0.61 & RGB \\
    S2  & 1.00 & 9.8445   & 5046$\pm$70  &  -0.75$\pm$0.1  & 2.39$\pm$0.20 & 64.52$\pm$1.94 & 30.31$\pm$0.58 & 4.04$\pm$0.05 & 304.20$\pm$3.04 & RC \\
    S3  & 1.60  & 9.3383   & 4830$\pm$70  &  0.0$\pm$0.1  & 2.92$\pm$0.20 & 25.57$\pm$0.77 & 105.01$\pm$1.83 & 8.66$\pm$0.05 & 70.80$\pm$0.71 & RGB \\
    S4  & 1.60  & 9.3461   & 4656$\pm$70  &  0.0$\pm$0.1  & 2.55$\pm$0.02 & 51.36$\pm$1.54 & 45.99$\pm$0.84 & 4.56$\pm$0.05 & 62.00$\pm$0.62 & RGB \\
    S5  & 1.60 & 9.3623  & 4769$\pm$70  &  0.0$\pm$0.1  & 2.54$\pm$0.20 & 58.40$\pm$1.75 & 43.97$\pm$0.81 & 4.60$\pm$0.05 & 268.30$\pm$2.68 & RC \\
    S6  & 2.35  & 8.9120   & 5003$\pm$70  &  0.0$\pm$0.1  & 2.85$\pm$0.20 & 51.41$\pm$1.54 & 86.79$\pm$1.54 & 6.86$\pm$0.05 & 251.20$\pm$2.51 & RC \\
    \hline 
    \end{tabular}
    \label{tab:artificial}
\end{table*}

In most cases, we recover the stellar masses and ages within the 68 per cent credible intervals. Using only \deltanu\ results in wider and more skewed PDFs (case \ref{item:first} in the plots), while adding \numax\ confines the solution in a much smaller region (cases \ref{item:second} and \ref{item:third}). When combining with \deltaP, the solution is tied better (case \ref{item:fifth}). In most cases, the combination of \deltanu\ and \numax\ provides narrower PDFs than \deltanu\ and \deltaP, which indicates that \deltaP\ does not constrain the solution as tightly as \numax\ (cases \ref{item:second} and \ref{item:fourth}) even for RC stars. As expected, adding more information as luminosity, narrows the searching ``area'' in the parameter space which provides the narrowest PDFs when all asteroseismic parameters and luminosity are combined (case \ref{item:seventh}). The usage of only \numax\ and luminosity (case \ref{item:eighth}) is very interesting, because it provides PDFs slightly narrower than using the typical combination of \deltanu\ and \numax\ or \deltanu\ and luminosity (case \ref{item:eleventh}), and it is similar to the \ref{item:fifth} and \ref{item:sixth} cases.  The lack of asteroseismic information (cases \ref{item:ninth}) worsens the situation, providing a significant larger error bar than most other cases, simply because of large uncertainties on gravity coming from spectroscopic analysis. The case \ref{item:tenth} results in PDFs very similar with case \ref{item:first}. 
This is to be expected: including \deltanu\ as a constraint (case \ref{item:first}) leads to a typical $\sigma(\log g) \simeq 0.02$ dex  (see also the discussion in \citealt{Morel14}, page 4), i.e., adding the spectroscopic $\log g$ ($\sigma(\log g) \simeq 0.2$ dex) as a constraint (case \ref{item:tenth}) has a negligible impact on the PDFs.

Finally, the prior on evolutionary stage does not change the shape of the PDFs in almost all cases, except for the RGB star S4. 
Regarding this case, it is interesting to note that S4 and S5 have similar \deltanu\ and \numax, but different \deltaP, that is, they are in a region of the \deltanu\ versus \numax\ diagram that is crossed by both RC and RGB evolutionary paths. In similar cases, not knowing the evolutionary stage causes the Bayesian code to cover all sections of the evolutionary paths, meaning that there is a large parameter space to cover,  which often causes the PDFs to become multi-peaked or spread for all possible solutions as the cases \ref{item:ninth} and \ref{item:tenth}. Further examples of this effect are given in figure 5 of \citet{rodrigues14}. Knowing the evolutionary stage, instead, limits the Bayesian code to weight just a fraction of the available evolutionary paths, hence limiting the parameter space to be explored and, occasionally, producing narrower PDFs. This is what happens for star S4, which, despite being a RGB star of 1.6~\Msun, happens to have asteroseismic parameters too similar to those the more long-lived RC stars of masses $\sim1.1$~\Msun.

Table~\ref{tab:rel_unc} presents the average relative mass and age uncertainties for RGB and RC stars, which summarizes well the qualitative description given above. Cases \ref{item:first} (very similar to case \ref{item:tenth}) and \ref{item:ninth} results the largest uncertainties: 17 and 12 per cent for RGB, and 8 and 11 for RC masses; up 70 and 40 per cent for RGB, and 22 and 31 for RC ages, respectively. From the traditional scaling relations (case \ref{item:second}) to the addition of period spacing and luminosity (case \ref{item:seventh}, the uncertainties can decrease from 8 to 3 per cent for RGB and 5 to 3 for RC masses; 29 to 10 per cent for RGB and 14 to 8 for RC ages. It is remarkable that we can also achieve a precision around 10 per cent on ages using \numax\ and luminosity (case \ref{item:eighth}), and 15 per cent using \deltanu\ and luminosity (case \ref{item:eleventh}).

\begin{table}
\scriptsize
\centering
\caption{Average relative uncertainties for each combination of input parameters for PARAM code as described in Section~\ref{sec:artificial}.}
\setlength{\tabcolsep}{4pt}
    \begin{tabular}{cccccc}
    \hline
     \multicolumn{2}{c}{Case}  &  \multicolumn{2}{c}{$<\sigma M/M>$} & \multicolumn{2}{c}{$<\sigma \text{Age}/\text{Age}>$} \\ \noalign{\smallskip}
     & &  RGB & RC & RGB & RC \\
    \hline
    \ref{item:first}  & \deltanu\ & 0.173  & 0.077 & 0.734    & 0.217  \\
    \ref{item:second} & \deltanu, \numax\ & 0.078  & 0.045 & 0.284    & 0.144   \\
    \ref{item:third} & \deltanu(SR), \numax\ & 0.061  & 0.047 & 0.220    & 0.146   \\
    \ref{item:fourth} & \deltanu, \deltaP\ & 0.109  & 0.052 & 0.336    & 0.181   \\
    \ref{item:fifth}  & \deltanu, \numax, \deltaP\  & 0.054  & 0.030 & 0.192    & 0.109   \\
    \ref{item:sixth}  & \deltanu, \deltaP, $L$  & 0.043  & 0.035 & 0.122    & 0.101   \\
    \ref{item:seventh} & \deltanu, \numax, \deltaP, $L$  & 0.034  & 0.025 & 0.097    & 0.075   \\
    \ref{item:eighth} & \numax, $L$ & 0.039  & 0.033 & 0.107    & 0.102   \\
    \ref{item:ninth}  & $\log{g}$, $L$ & 0.124  & 0.108 & 0.427   & 0.310  \\ 
    \ref{item:tenth}  &\deltanu\, $\log{g}$ & 0.173 & 0.077 & 0.727 & 0.215 \\
    \ref{item:eleventh}  &\deltanu\, $L$ & 0.052 & 0.046  & 0.143 & 0.146  \\
\hline 
\end{tabular}
\label{tab:rel_unc}
\end{table}

Average relative differences between masses are $\leq$ 1 per cent for cases \ref{item:fifth}, \ref{item:sixth}, \ref{item:seventh}, and \ref{item:eighth}, around 1 per cent for cases \ref{item:second} and \ref{item:eleventh}, $\sim6$ per cent for case \ref{item:third}, and greater than 6 per cent for cases \ref{item:first}, \ref{item:ninth}, and \ref{item:tenth}. Regarding ages, relative absolute differences are lesser than 5 per cent for cases \ref{item:fifth}, \ref{item:sixth}, \ref{item:seventh}, \ref{item:eighth}, and \ref{item:eleventh}, around 10 per cent when using \deltanu\ and \numax, $\sim20$ per cent when using \deltanu(SR) and \numax, and greater than 40 per cent for cases \ref{item:first}, \ref{item:ninth}, and \ref{item:tenth}.

We also applied mass loss on the models. Figure~\ref{fig:pdf_m_age_eta} shows the resulting mass and age PDFs for stars S2 and S5 with the efficiency parameter $\eta=0.2$ (cyan colors) and $\eta=0.4$ (white colors). For the cases \ref{item:fifth}, \ref{item:sixth}, \ref{item:seventh}, \ref{item:eighth}, and \ref{item:eleventh}, a mass loss with efficiency $\eta=0.4$ produces differences on masses of $\sim$1 per cent, while on ages, may be greater than 47 per cent for S2 and than 18 per cent for S5. The small difference in masses results from the fact that, in these cases, mass values follow almost directly from the observables -- roughly speaking, they represent the mass of the tracks that pass closer to the observed parameters. As well known, red giant stars quickly lose memory of their initial masses and follow evolutionary tracks which are primarily just a function of their actual mass and surface chemical composition. So their derived masses will be almost the same, irrespective of the mass loss employed to compute previous evolutionary stages. But the value of $\eta$ will affect the relationship between the actual masses and the initial ones at the main sequence, which are those that determine the stellar age. For instance, S2 have nearly the same actual mass (very close to 1~\Msun) for both $\eta=0.2$ and $\eta=0.4$ cases, but this actual mass can derive from a star of initial mass close to 1.075~\Msun\ in the case of $\eta=0.2$, or from a star of initial mass close to 1.15~\Msun\ in the case of $\eta=0.4$. This $\sim13$ per cent difference in the initial, main sequence mass is enough to explain the $\sim47$ per cent difference in the derived ages of S2. More in general, this large dependence of the derived ages on the assumed efficiency of mass loss, warns against trusting on the ages of RC stars.


\subsection{NGC~6819}
\label{sec:ngc6819}

The previous section demonstrates that it is possible to recover, generally within the expected 68~per cent (1$\sigma$) credible interval expected from observational errors, the masses and ages of artificial stars. It is not granted that a similar level of accuracy will be obtained in the analysis of real data. Star clusters, whose members are all expected to be at the same distance and share a common initial chemical composition and age, offer one of the few possible ways to actually verify this. Only four clusters have been observed in the {\em Kepler} field \citep{gilliland10}, and among these NGC~6819 represents the best case study, owing to its brightness, its  near-solar metallicity (for which stellar models are expected to be better calibrated) and the large numbers of stars in {\em Kepler} database. NGC~6791 has even more giants observed by {\em Kepler};  however its super-solar metallicity, the uncertainty about its initial helium content, and larger age -- causing a non-negligible mass loss before the RC stage -- makes any comparison with evolutionary models more complicated.
 
\citet{handberg16} reanalysed the raw {\it Kepler} data of the stars in the open cluster NGC~6819 and extracted individual frequencies, heights, and linewidths for several oscillation modes. They also derived the average seismic parameters and stellar properties for $\sim$50 red giant stars based on targets of \citet{stello11_1}. Effective temperatures were computed based on $V-K_s$ colours with bolometric correction and intrinsic colour tables from \citet{casagrande_vandenberg14}, and adopting a reddening of $E(B-V)=0.15$~mag. They derived masses and radii using scaling relations, and computed apparent distance moduli using bolometric corrections from  \citet{casagrande_vandenberg14}. The authors also applied an empirical correction of 2.54 per cent to the \deltanu\ of RGB stars, thus making the mean distance of RGB and RC stars to become identical. As we based the definition of the average \deltanu\ for MESA models similar to the one used in \citet{handberg16}'s work, we adopted their values for the global seismic (\deltanu\ and \numax) and spectroscopic (\Teff) parameters.  
We verified that their \Teff\ scale is just $\sim\!57$~K cooler than the spectroscopic measurements from the APOGEE Data Release 12 \citep{alam15}. The metallicity adopted was $\feh=0.02\pm0.10$ dex for all stars.
We also adopted period spacing values from \citet{Vrard2016}, who automatically measured \deltaP\ for more than 6000 stars observed by {\it Kepler}.  
In order to derive distances and extinctions in the $V$-band ($A_V$), we also used the following apparent magnitudes: SDSS $griz$ measured by the KIC team \citep{brown11} and corrected by \citet{pinsonneault12}; $JHK_s$ from 2MASS \citep{cutri03,skrutskie06}; and $W1$ and $W2$ from WISE \citep{wright10}.

We computed stellar properties for 52 stars that have \Teff, \feh, \deltanu, and \numax\ available using case \ref{item:second} and \ref{item:third}; and for 20 stars that have also \deltaP\ measurements using case \ref{item:fifth}.  Table~\ref{tab:cluster_rel_unc} presents the average relative uncertainties on masses and ages for these stars. These average uncertainties are slightly smaller than the ones from our test with artificial stars in the previous section.

\begin{table}
\scriptsize
\centering
\caption{Average relative uncertainties on masses and ages for stars in NGC~6819 using the combination of input parameters \ref{item:second}, \ref{item:third}, and \ref{item:fifth} for PARAM code.}
\setlength{\tabcolsep}{4pt}
    \begin{tabular}{cccccc}
    \hline
     \multicolumn{2}{c}{Case}  &  \multicolumn{2}{c}{$<\sigma M/M>$} & \multicolumn{2}{c}{$<\sigma \text{Age}/\text{Age}>$} \\ \noalign{\smallskip}
     & &  RGB & RC & RGB & RC \\
    \hline
    \ref{item:second} & \deltanu, \numax\ & 0.057 & 0.026 & 0.210    & 0.100   \\
    \ref{item:third} & \deltanu(SR), \numax\ & 0.044  & 0.026 &  0.161    & 0.102   \\
    \ref{item:fifth} & \deltanu, \numax, \deltaP\ &  0.013 & 0.021 & 0.050 & 0.077 \\
\hline 
\end{tabular}
\label{tab:cluster_rel_unc}
\end{table}

\begin{figure}
\resizebox{\hsize}{!}{\includegraphics{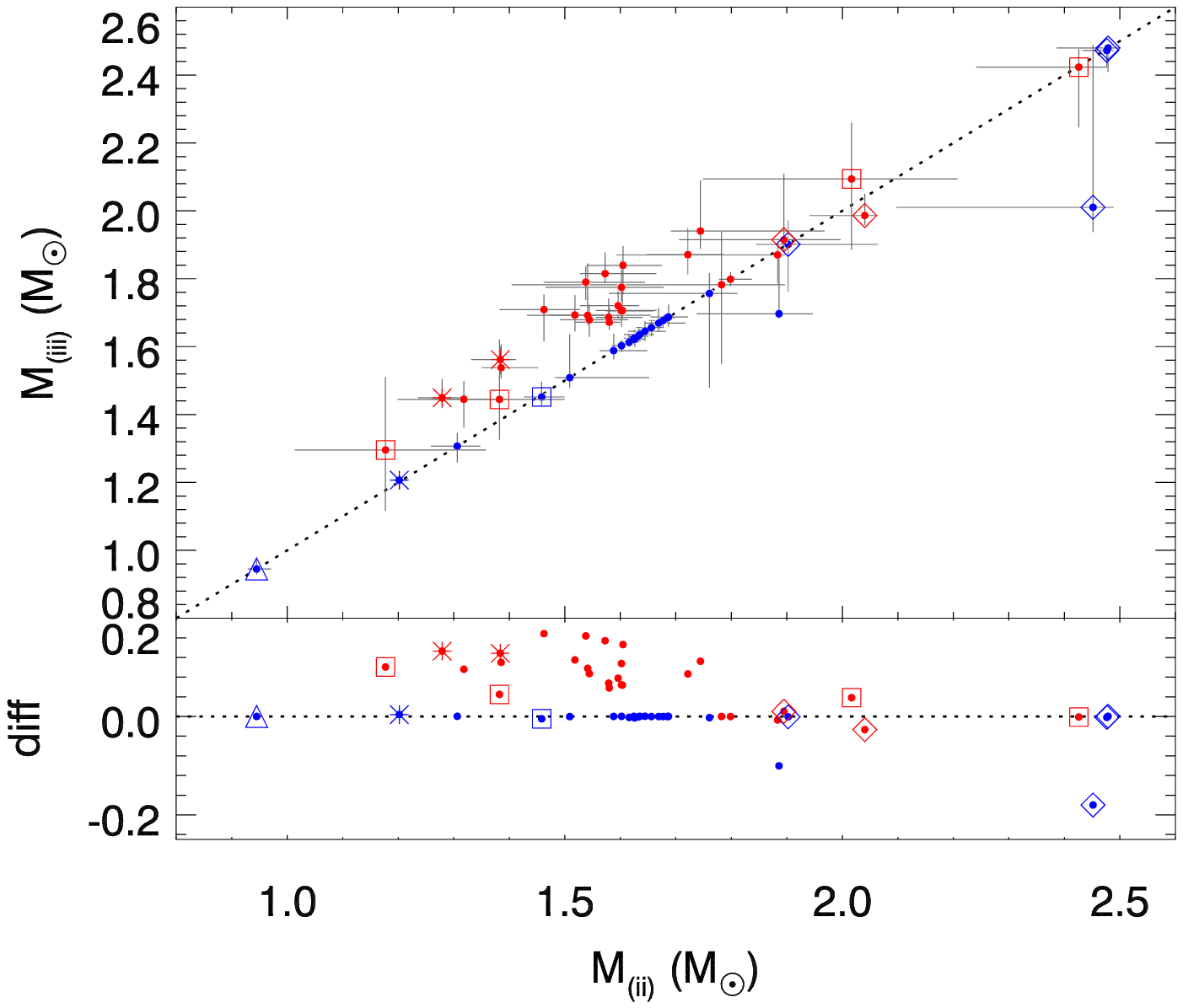}}
\resizebox{\hsize}{!}{\includegraphics{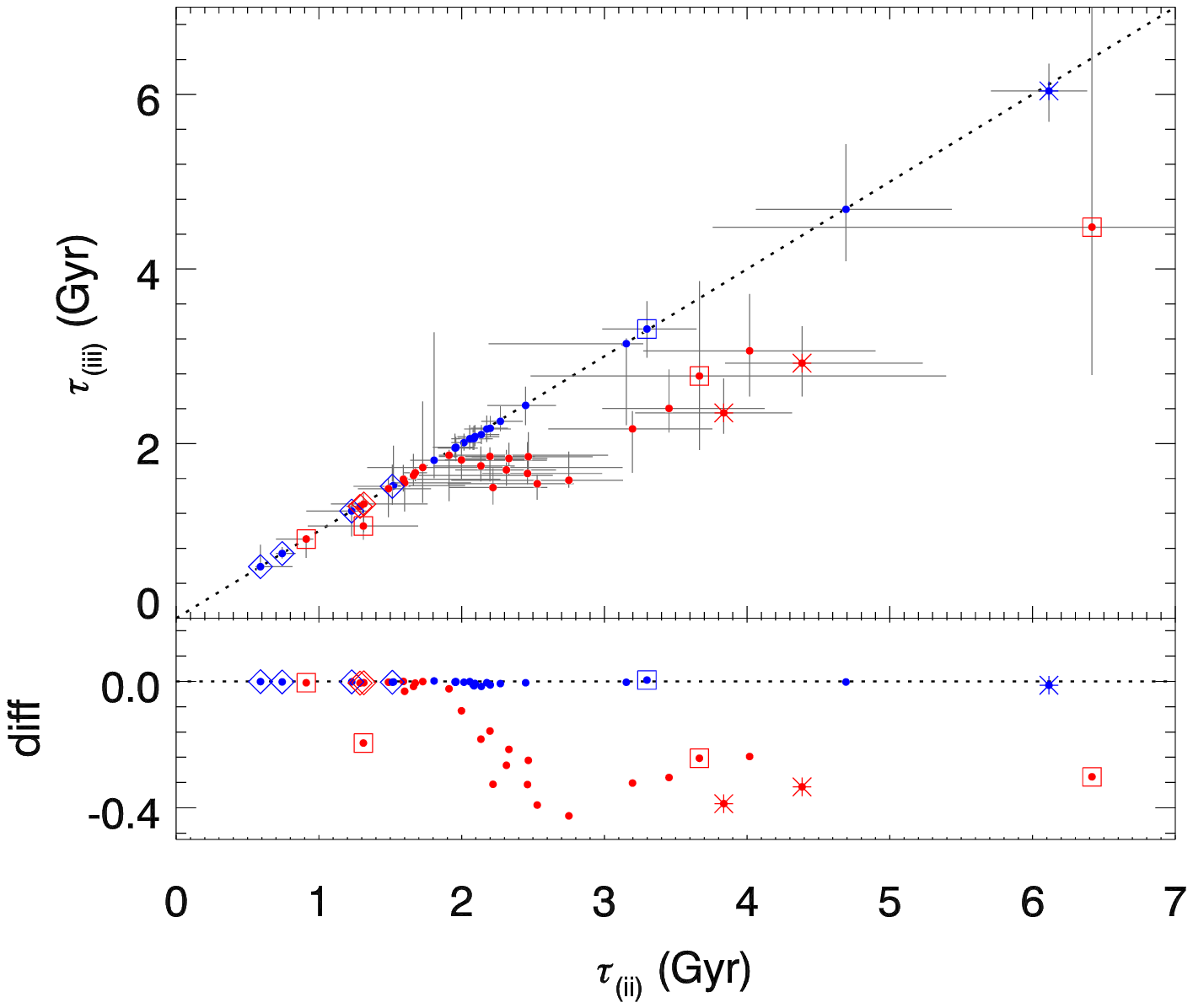}}
\caption{Comparison between masses (top panel) and ages (bottom) estimated with case \ref{item:second} versus case \ref{item:third}. The bottom panel excludes KIC~4937011 (Li-rich low mass RC) that has an estimated age in both cases of $\sim13.8$~Gyr. Sub-panels show relative differences. Dotted black lines are the identity line. The blue and red colors represent RC and RGB stars, respectively. Different symbols are peculiar stars that were discussed in details in \citet{handberg16} -- asterisks are stars classified as non-member; diamonds: stars classified as over-massive; squares: uncertain cases; triangle: Li-rich low mass RC (KIC 4937011).}
\label{fig:cluster_mass_age}
\end{figure}

Figure~\ref{fig:cluster_mass_age} shows the masses and ages derived using PARAM with case \ref{item:second} and \ref{item:third} as observational input. The blue and red colors represent RC and RGB stars, respectively. The median and mean relative differences between stellar properties are presented in Table~\ref{tab:reldiff_IF_SR}. The RGB stars have masses $\sim 8$ per cent greater when using \deltanu\ scaling relation, while many RC stars present no difference and only few of them have smaller masses ($\approx 2$ per cent). The mass differences reflects RGB stars being on average $\sim 18$ per cent younger and no significant differences on RC stars. The $\sim 5$ per cent difference on RGB radii reflects on the same difference on distances.

\begin{table}
\scriptsize
    \centering
    \caption{Median and mean relative (and absolute) differences between properties estimated using case~\ref{item:second} and \ref{item:third} for RGB and RC stars from NGC~6819.}
    \label{tab:reldiff_IF_SR}
    \begin{tabular}{ccccc}
    \hline
    \multirow{2}{*}{properties} & \multicolumn{2}{c}{RGB} & \multicolumn{2}{c}{RC} \\
                                & median     & mean       & median     & mean      \\ \hline
    masses                      &  0.088    & 0.079   & 0.000    & -0.012   \\ 
    ages                        & -0.195    & -0.180    & -0.002 &  -0.004   \\
    radii                    & 0.048    & 0.043     & 0.000 &  -0.007   \\ 
    $A_V$                        & 0.005     & 0.031    &  0.001     &  0.001   \\ 
    distances                   &  0.047     & 0.045     &  -0.001    & -0.006  \\ \hline
\end{tabular}
\end{table}

\begin{figure}
    \resizebox{\hsize}{!}{\includegraphics{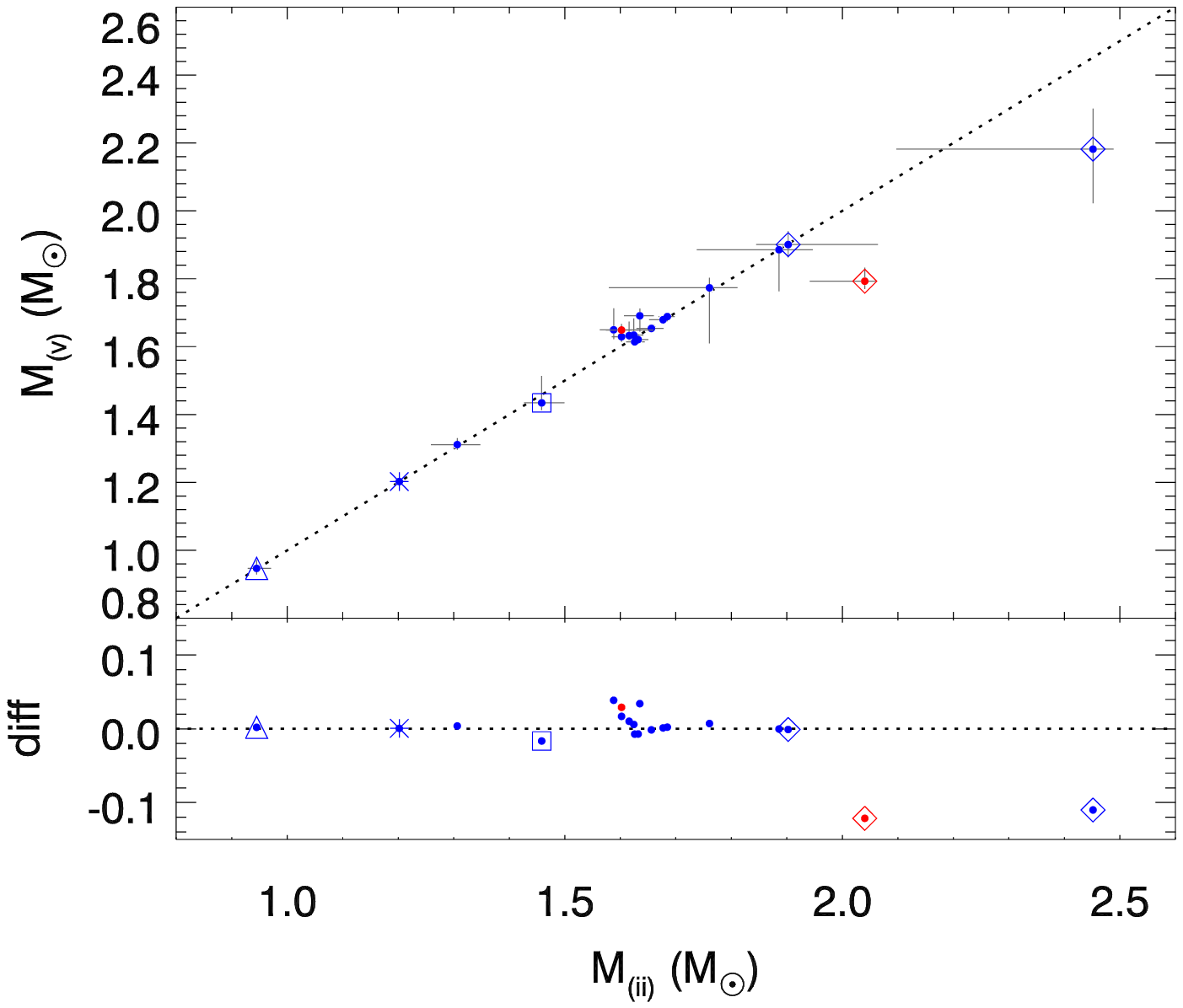}}
    \resizebox{\hsize}{!}{\includegraphics{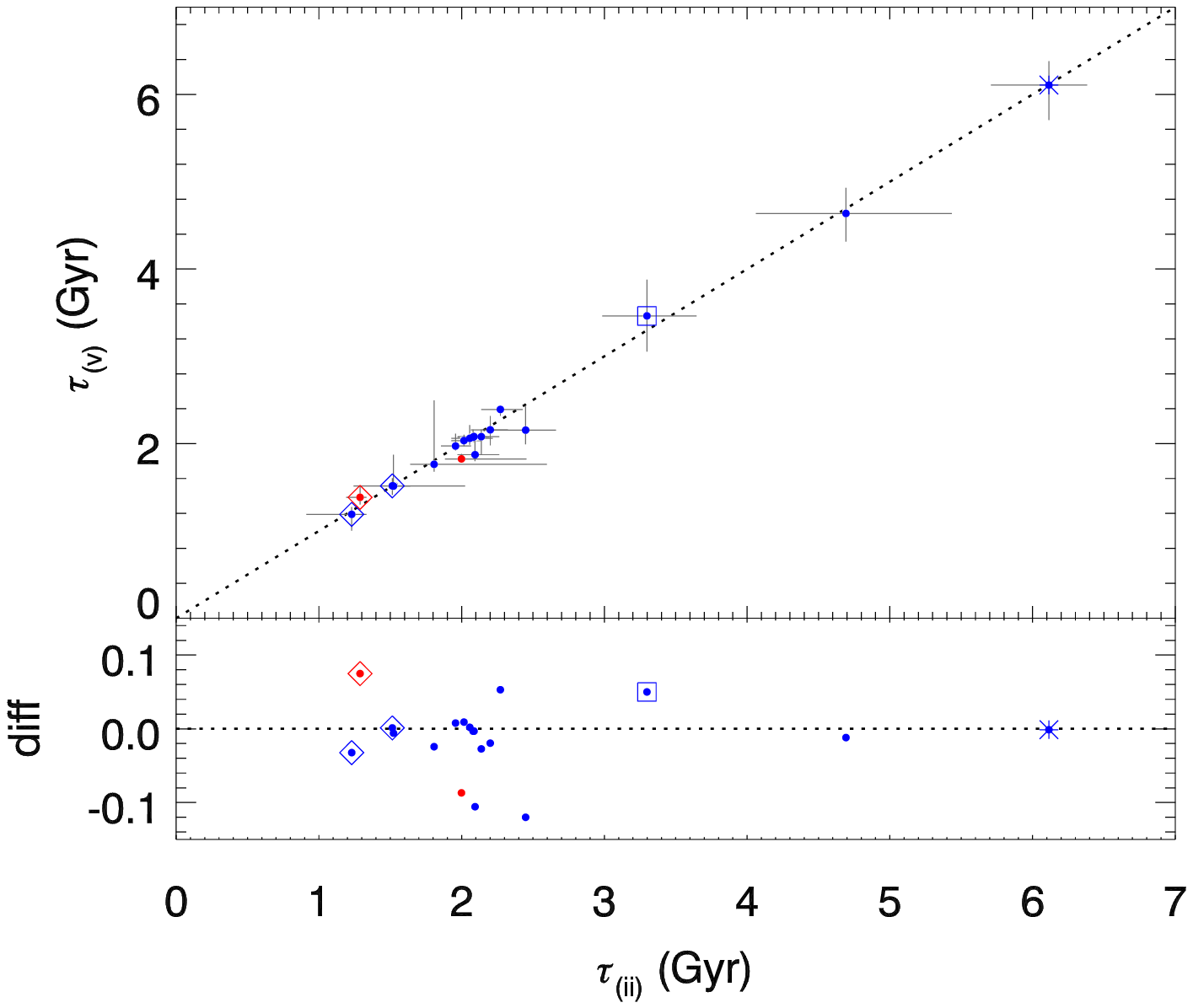}}
    \caption{Same as Fig.~\ref{fig:cluster_mass_age}, but with case \ref{item:second} versus case \ref{item:fifth}.}
    \label{fig:cluster_mass_age_v}
\end{figure}

Figure~\ref{fig:cluster_mass_age_v} shows the masses and ages derived using case \ref{item:second} and \ref{item:fifth} as observational input. The average relative uncertainties are much smaller for RGB stars when adding \deltaP\ as an observational constraint (see Table~\ref{tab:cluster_rel_unc}). The agreement on masses is very good, except for massive stars, when masses are around the upper mass limit of our grid (2.50~\Msun). Two over-massive stars result $\sim 10$ per cent less massive when adding $\deltaP$ (KIC~5024476 and 5112361). The ages also present a good agreement inside the error bars, although with a dispersion of $\sim\!5$ per cent. 

\begin{figure}
    \resizebox{\hsize}{!}{\includegraphics{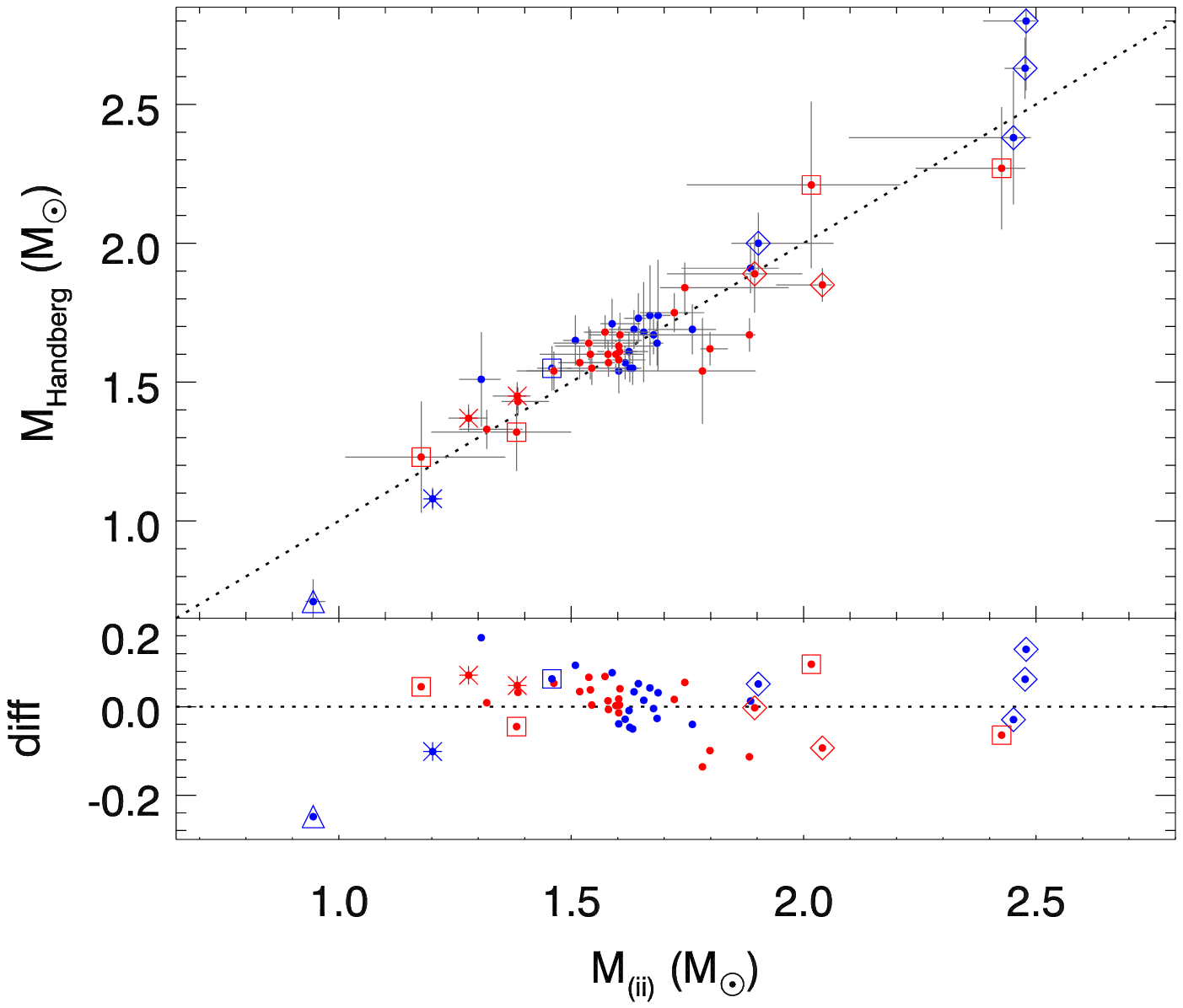}}
    \resizebox{\hsize}{!}{\includegraphics{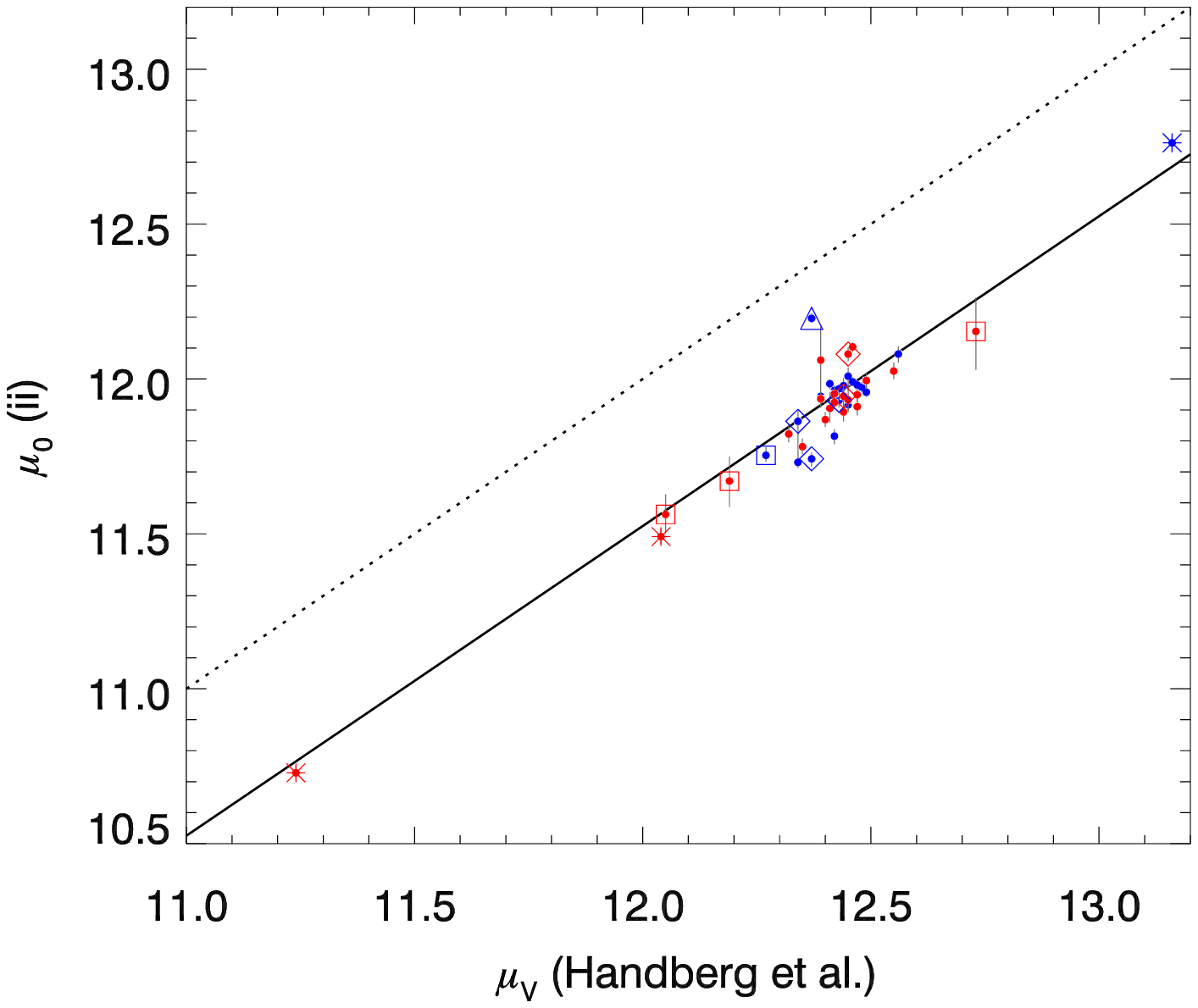}}
    \caption{Comparison between masses (top panel) and distance moduli (bottom) estimated with case \ref{item:second} and from \citet{handberg16}. Dotted black lines are the identity line. The blue and red colors represent RC and RGB stars, respectively. Different symbols are the same as Fig.~\ref{fig:cluster_mass_age}. The solid black line in the bottom panel shows the agreement between our distance with the distance in the V-band, representing a measurement of the extinction.}
    \label{fig:cluster_mass_hand}
\end{figure}

The top panel of figure~\ref{fig:cluster_mass_hand} shows the comparison between masses estimated with case \ref{item:second} versus masses from \citet{handberg16}. The masses have a good agreement with a dispersion of $\sim 7$ per cent, showing that the proposed correction of 2.54 per cent on \deltanu\ for RGB stars in \citet{handberg16} compensates the deviations when using \deltanu\ scaling. The authors also discussed in details some stars that seem to experience {\it non-standard} evolution based on their masses and distances estimations and on membership classification based on radial velocity and proper motion study by \citet{milliman14}. These stars are represented with different symbols in all figures of this section: asterisks -- non-member stars (KIC~4937257, 5024043, 5023889); diamonds -- stars classified as overmassive (KIC~5024272, 5023953, 5024476, 5024414, 5112880, 5112361); squares -- uncertain cases (KIC~5112974, 5113061, 5112786, 4937770, 4937775); triangle -- Li--rich low mass RC (KIC 4937011). A similar detailed description star by star is not the scope of the present paper, however the peculiarities of these stars should be kept on mind when deriving their stellar and the cluster properties. Some of the over-massive stars do not have a good agreement, because of the upper mass limit of our grid of models ($2.5~\Msun$). Taking into account only single member stars, the mean masses of RGB and RC stars using case \ref{item:second} are  $1.61\pm0.04$~\Msun\ and $1.62\pm0.03$~\Msun, which also agree with the ones found in \citet{handberg16} and \citet{miglio12_2}.

\begin{figure}
    \resizebox{\hsize}{!}{\includegraphics{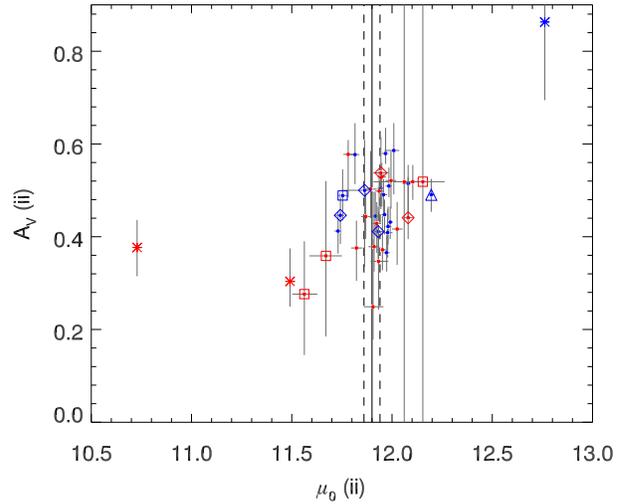}}
    \caption{Extinction versus distance moduli estimated with case \ref{item:second}. The blue and red colors represent RC and RGB stars, respectively. Solid and dashed black lines are the mean and its uncertainty of distance moduli computed taking into account all stars, except for the ones classified as non-member (asterisks) by \citet{handberg16}. Different symbols are the same as Fig.~\ref{fig:cluster_mass_age}.}
    \label{fig:cluster_mu0_av}
\end{figure}

The bottom panel of figure~\ref{fig:cluster_mass_hand} shows the comparison between distance moduli estimated with case \ref{item:second} versus distance moduli in the $V$-band estimated in \citet{handberg16}. The solid line represents the linear regression $\mu_0 = \mu_V(\text{Handberg}) - A_V$, which results $A_V=0.475\pm0.003$~mag that is in a good agreement with the average extinction for the cluster (see Fig.~\ref{fig:cluster_mu0_av}). Our method estimates the extinction star-to-star and it varies significantly in the range $A_V=[0.3,0.7]$ for the stars in the cluster. This seems to be in agreement with \citet{platais16} that shown a substantial differential reddening in this cluster with the maximum being $\Delta E(B-V)=0.06$ mag, what implies extinctions in the $V$-band in the same range that we found. Extinctions and distance moduli estimated using case \ref{item:second} are presented in Figure~\ref{fig:cluster_mu0_av}. The average uncertainties on extinctions and distance moduli are 0.1~mag and 0.03~mag ($<2$ per cent on distances), respectively. We derived the distance for the cluster by computing the mean distances, $\mu_0=11.90\pm0.04$~mag with a dispersion of 0.23~mag (solid and dashed black lines in Figure~\ref{fig:cluster_mu0_av}), excluding stars classified as non-member (asterisks) by \citet{handberg16}. This value compares well with distance moduli measured for eclipsing binaries, $\mu_0=12.07\pm0.07$~mag \citep{jeffries13}. 

\begin{figure}
    \resizebox{\hsize}{!}{\includegraphics{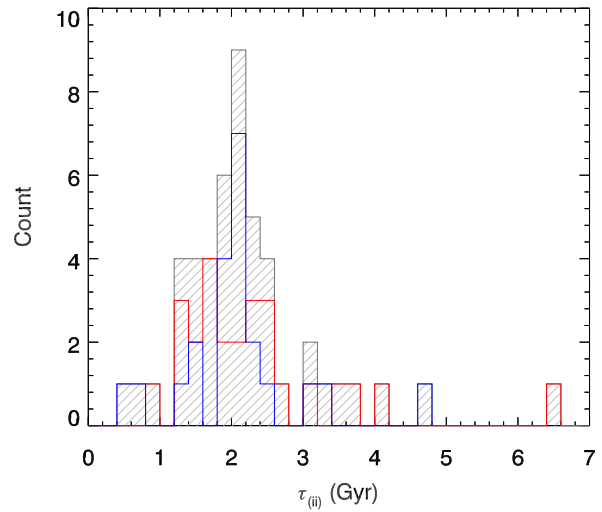}}
    \caption{Histogram of ages estimated using case \ref{item:second}. The gray line represents all stars, except the ones classified as non-members stars and KIC~4937011 that has $\sim 13.8$ Gyr. Red and blue lines represent the ages of RGB and RC stars.}
\label{fig:cluster_hist_age}
\end{figure}

Figure~\ref{fig:cluster_hist_age} shows the histogram of the age estimated using case \ref{item:second}. The gray line represents the histogram of all stars, except the 3 stars classified as non-member and the star KIC~4937011 that likely experienced very high mass-loss during its evolution (see discussion in \citealt{handberg16}). Red and blue lines represent the ages of RGB and RC stars. The mean age by the gray histogram is $2.22\pm0.15$~Gyr with a dispersion of 1.01~Gyr, that agrees with the age estimated by fitting isochrones to the cluster CMDs by \citet{brewer16} ($2.21 \pm0.10 \pm 0.20$ Gyr). Taking into account only stars classified as single members (31 stars), i.e. excluding stars that are binary members, single members flagged as over-under massive and with uncertain parameters classified according to \citet{handberg16}, the mean age results $2.25\pm0.12$~Gyr with a dispersion of 0.64~Gyr.
Importantly, RGB and RC apparently share the same age distribution, i.e. there is no evidence of systematic differences in the ages of the two groups of stars. This result reflects taking into consideration the deviations from scaling relations, which are quite relevant for RGB stars but smaller for the RC. Adding \deltaP\ (case \ref{item:fifth}), the mean age is $2.12\pm0.19$~Gyr with a dispersion of 0.79~Gyr, excluding the star KIC~4937011 and also the one classified as non-member KIC~4937257 (triangle and asterisk symbols in Figure~\ref{fig:cluster_mass_age_v}). For this case, there are 13 stars classified as single member according to \citet{handberg16}, whose mean age is $2.18\pm0.20$~Gyr with a dispersion of 0.73~Gyr.
In the case with \deltanu\ scaling (case \ref{item:third}) the mean age is $1.95\pm0.11$~Gyr (dispersion of 0.78~Gyr, computed also excluding the 3 stars classified as non-member and the star KIC~4937011), 12 per cent younger than using \deltanu\ from models.

\begin{figure}
    \resizebox{\hsize}{!}{\includegraphics{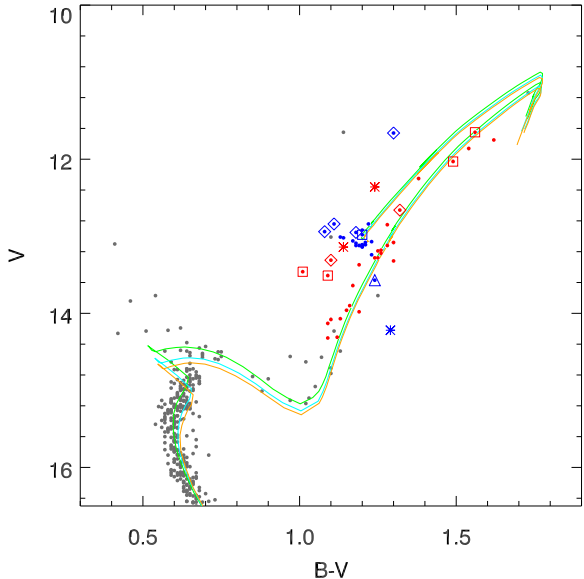}}
    \caption{CMD for the cluster stars with membership probability $\geq 90$ per cent according to radial velocity by \citet{hole09} (gray dots). The blue and red colors represent RC and RGB stars, respectively. Different symbols are the same as Fig.~\ref{fig:cluster_mass_age}. The green, cyan, and orange lines are MESA isochrones with ages 2.0, 2.2, and 2.3~Gyr, using $\mu_0=11.90$~mag and $E(B-V)=0.14$~mag.}
    \label{fig:cluster_cmd}
\end{figure}

Figure~\ref{fig:cluster_cmd} shows the color-magnitude diagram (CMD) for the cluster stars with membership probability $\geq 90$ per cent according to radial velocity by \citet{hole09} (gray dots). The red and blue symbols are the stars analysed in the present work. There is a significant dispersion on the RGB and RC, but still our isochrones match well the photometry. This points to a significant consistency between the ages of evolved stars derived from asteroseismology, and the CMD-fitting age which would be derived from the photometry. This particular result, however, should not be generalised, since it applies only to the specific set of stellar models and cluster data that has been used here.

Another important aspect, however, is that the ages derived for cluster stars turn out to present a  larger scatter than expected. If we assume that all cluster stars really have the same age, their mean standard deviation implies that the final errors in the ages are of roughly 46 per cent, which is a factor of 2 larger than the individual age uncertainties for the case \ref{item:second} (see Table~\ref{tab:cluster_rel_unc}).

The scatter is reduced when excluding from the sample stars that are binary members, single members flagged as over-under massive and with uncertain parameters classified according to \citet{handberg16}. In this case the scatter  (28 per cent) is higher, but comparable with, the expected uncertainty (21 per cent).

At present, the origin of this increased age dispersion is not clear. We note however that the NGC~6819 giants are also dispersed around the best-age isochrones in the CMD. The magnitude of this dispersion is not simply attributable to differential reddening or photometric errors \citep{hole09,milliman14,brewer16}. Therefore, it is possible that it reflects some physical process acting in the individual cluster stars, rather than a failure in the method.

We also notice that in the cluster CMD (Fig.~\ref{fig:cluster_cmd}) the main sequence turn-off is well-defined and the comparison with isochrones appear to rule out internal age spreads larger than $\sim0.2$~Gyr. Even larger age spreads have been suggested to explain the very extended (and sometimes bimodal) main sequence turn-offs observed in some very massive star clusters in the Magellanic Clouds \citep[][and references therein]{goudfrooij15}. However, there is no evidence of a similar feature occurring in the photometry of NGC~6819.


\section{Discussion and conclusions}
\label{sec:close}

Our main conclusions are:
\begin{itemize}
\item It is possible to implement the asteroseismic quantities \deltanu\ and \deltaP, computed along detailed grids of stellar evolutionary tracks, into the usual Bayesian or grid-based methods of parameter estimation for asteroseismic targets. We perform such an implementation in the PARAM code. It will be soon become available for public use through the web interface \url{http://stev.oapd.inaf.it/param}.
\item Tests with synthetic data reveal that masses and ages can be determined with typical precision of 5 and 19 per cent, if precise global seismic parameters (\deltanu, \numax, \deltaP) are available. Adding luminosity these values can decrease to 3 and 10 per cent, respectively. 
\item Combining the luminosity expected from the end-of-mission Gaia  parallaxes with \deltanu,  enables us to infer masses (ages) to $\sim 5$ per cent ($\sim 15$ per cent)  independently from the \numax\ scaling relation, which is still lacking a detailed theoretical understanding (but see \citealt{belkacem11}). A similar precision on mass and age is also expected when combining luminosity and \numax: this will be particularly relevant for stars where data are not of sufficient quality/duration to enable a robust measurement of \deltanu. Stringent tests of the accuracy of the \numax\ scaling relation \citep[as in][]{coelho15} are therefore of great relevance in this context.
\item Any estimate based on asteroseismic parameters is at least a factor 4 more precise than those based on spectroscopic parameters alone.
\item The application of these methods to NGC~6819 giants produces mean age of $2.22\pm0.15$~Gyr, distance $\mu_0=11.90\pm0.04$~mag, and extinctions $A_V\approx 0.475\pm0.003$~mag. All these values are in agreement with estimates derived from photometry alone, via isochrone fitting.
\item Despite these encouraging results, the application of the method to NGC~6819 stars also reveals a few caveats and far-from-negligible complications. Even after removing some evident outliers (likely non members) from the analyses, the age dispersion of NGC~6819 stars turns out to be appreciable, with the $\tau=2.22\pm0.15$~Gyr with a dispersion of 1.01~Gyr, implying a $\sim46$~per cent error on individual ages (or $\sim28$~per cent taking into account only single members and removing over-massive stars indentified in \citealt{handberg16}). The mean age value is compatible with those determined with independent methods (e.g.\ the $\tau=2.21\pm0.10\pm0.20$~Gyr from isochrone fitting).
\end{itemize}

The result of a large age dispersion for NGC~6819 stars is no doubt surprising, given the smaller typical errors found during our tests with artificial data. Since asteroseismology is now widely regarded as the key to derive precise ages for large samples of field giants distributed widely across the Galaxy, this is surely a point that has to be understood:  any uncertainty or systematics affecting the NGC~6819 stars will also affect the analyses of the field giants observed by asteroseismic missions. 

We could point out that, on the one hand, a clear source of bias in age is the presence of over/under-massive stars which are likely to be the product of binary evolution. Additionally, even restricting ourselves to RGB stars and weeding out clear over/under massive stars we are left with an age/mass spread which is larger than expected (28 per cent compared to 21 per cent). Grid-based modelling increases the significance of this spread, compared to the results presented in \citet{handberg16}.

Whether this spread is an effect specific to the age-metallicity of NGC~6819, is yet to be determined. 
Previous works on NGC~6791 and M~67, for instance, have not reported on a significant spread in mass/age  of their asteroseismic targets \citep[]{basu11,miglio12_2,corsaro12,stello16}. These three clusters are different in many aspects, with NGC~6791 being the most atypical one given its very high metallicity. Apart form this obvious difference, in both NGC~6791 and M~67 the evolved stars have masses smaller than 1.4~\Msun, and were of spectral type mid/late-F or G -- hence slow rotators -- while in their main sequence. 
In NGC~6819 the evolved stars have masses high enough to be ``retired A-stars'', which includes the possibility of having been fast rotators before becoming giants. This is a difference that could, at least partially, be influencing our results. Indeed, rotation during the main sequence is able to change the stellar core masses, chemical profile, and main sequence lifetimes \citep{eggenberger10, lagarde16}. A spread in rotational velocities among coeval stars might then cause the spread in the properties of the red giants, which might not be captured in our grids of non-rotating stellar models. The possible impact of rotation in the grid-based and Bayesian methods, has still to be investigated.

On the other hand, this $\sim46$~per cent uncertainty is comparable to the 0.2~dex uncertainties that are obtained for the ages of giants with precise spectroscopic data and {\em Hipparcos} parallax uncertainties smaller than 10~\% \citep{feuillet16}, which refer to stars within 100~pc of the Sun. In this sense, our results confirm that asteroseismic data offer the best prospects to derive astrophysically-useful ages for individual, distant stars.


\section*{Acknowledgments}
We thank the anonymous referee for his/her useful comments.
We acknowledge the support from the PRIN INAF 2014 -- CRA 1.05.01.94.05. TSR acknowledges support from CNPq-Brazil. JM and MT acknowledge support from the ERC Consolidator Grant funding scheme ({\em project STARKEY}, G.A. n. 615604). AM acknowledges the support of the UK Science and Technology Facilities Council (STFC). Funding for the Stellar Astrophysics Centre is provided by The Danish National Research Foundation (Grant agreement no.: DNRF106).
%
\bibliographystyle{mn2e} 
\bibliography{main} 
%
%
\label{lastpage}
\end{document}